%% file: paper.tex
\documentclass{egpubl}
\JournalPaper         % uncomment for final version of Journal Paper
\usepackage[T1]{fontenc}
\usepackage{dfadobe}  

\usepackage{cite}  % comment out for biblatex with backend=biber
\BibtexOrBiblatex
\electronicVersion
\PrintedOrElectronic
\ifpdf \usepackage[pdftex]{graphicx} \pdfcompresslevel=9
\else \usepackage[dvips]{graphicx} \fi

\usepackage{egweblnk} 
\input{macro}

\title[ClipFlip : Multi-view Clipart Design]%
      {ClipFlip : Multi-view Clipart Design}

\author[I-Chao Shen et al.]
{\parbox{\textwidth}{\centering I-Chao Shen$^{1}$\orcid{0000-0003-4201-3793} ~ Kuan-Hung Liu$^{1}$ ~ Li-Wen Su$^{1}$ ~ Yu-Ting Wu$^{1}$ ~ Bing-Yu Chen$^{2}$
        }
        \\
{\parbox{\textwidth}{\centering $^1$ \{jdily, lkh94, susan31213, kevincosner\}@cmlab.csie.ntu.edu.tw, National Taiwan University, Taiwan \\
$^2$ robin@ntu.edu.tw, National Taiwan University, Taiwan
       }
}
}

\begin{document}

\teaser{
 \includegraphics[width=\linewidth]{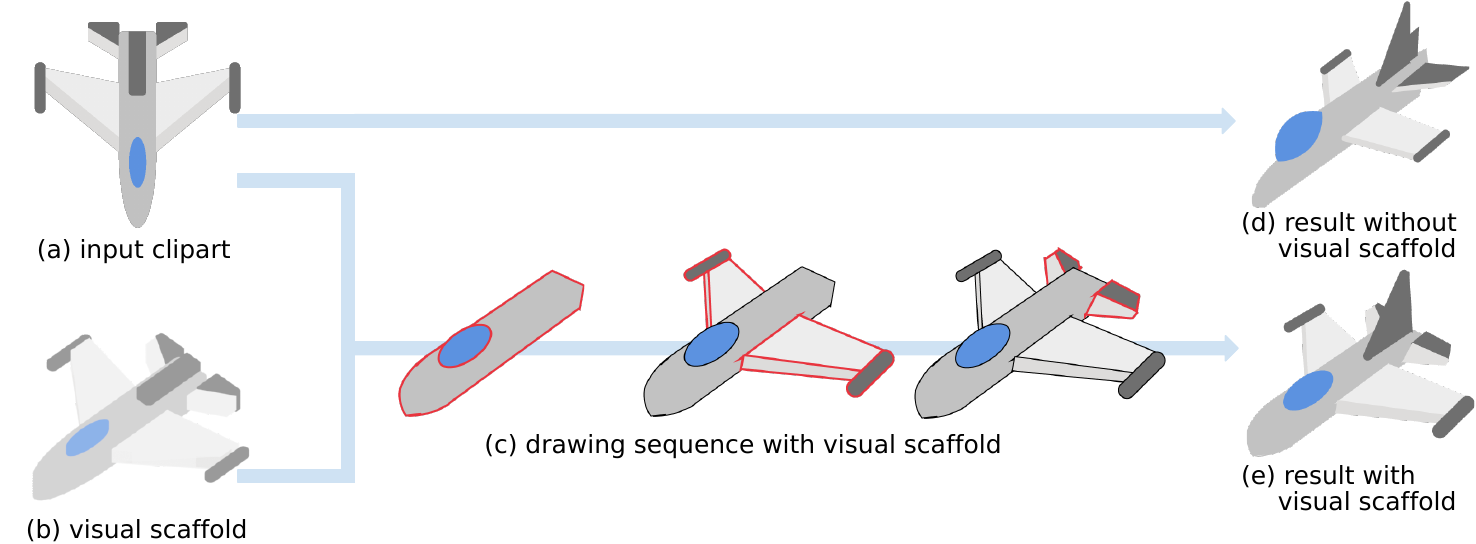}
 \centering
  \caption{Given the input clipart (a), Our system automatically generates a visual scaffold (b) by rendering it from a style consistent 3D shape generated by a user-assisted curve extrusion method.
  The user can design the clipart from the desired viewpoint by following the visual scaffold and draw each part step-by-step (c).
  We highlight the parts are being drawn at each step in red.
  Finally, the user can design better clipart (e) from the desired viewpoint in a  shorter period than designing it without the provided visual scaffold (d).
  }
\label{fig:teaser}
}

\maketitle

%%% content
\input{abstract}
\input{introduction}
\input{relatedwork}
\input{method}
\input{userinterface}
\input{result}
\input{conclusion}
\input{ack}
% bibtex
\bibliographystyle{eg-alpha-doi}  
\bibliography{paper.bib}        

% biblatex with biber
% \printbibliography       
\end{document}

%% file: macro.tex
\usepackage{enumitem}
\usepackage{amsmath}
\usepackage{amsfonts}
\usepackage{multirow}
\usepackage{tabu}
\usepackage{makecell}
\usepackage{soul}
\usepackage{color}
\usepackage{booktabs}
\usepackage{ifthen}
\usepackage{hyperref}
\usepackage{subcaption}
\usepackage{bbold}
\usepackage{wrapfig}
\usepackage[linesnumbered,ruled,vlined]{algorithm2e}
\usepackage{mathtools}
\usepackage{xcolor}% http://ctan.org/pkg/xcolor
\usepackage[normalem]{ulem} % for sout
\definecolor{gray}{rgb}{0.5,0.5,0.5}
\definecolor{green}{rgb}{0, 0.6, 0}
\definecolor{orange}{rgb}{1, 0.5, 0}
\definecolor{mahogany}{rgb}{0.75, 0.25, 0.0}
\definecolor{purple}{rgb}{0.6, 0, 0.6}
\definecolor{darkgreen}{rgb}{0, 0.3, 0}
\definecolor{orange}{rgb}{1, 0.5, 0.}

\newcommand{\cgfhl}[1]{#1}
\newcommand{\minorhl}[1]{#1}
\newcommand{\minorhll}[1]{#1}

\newcommand{\ignore}[1]{}
\newcommand{\none}[1]{}
\newcommand{\com}[1]{}
\newcommand{\etal}{{\textit{et~al.}}}
\newcommand{\ie}{i.e.,}
\newcommand{\eg}{e.g.,}
\newcommand{\figname}{Figure}

\newcommand{\secname}{Section}

\newcommand{\eqname}{Eq.}

%% file: abstract.tex
\begin{abstract}
We present an assistive system for clipart design by providing visual scaffolds from the unseen viewpoints.
Inspired by the artists' creation process, our system constructs the visual scaffold by first synthesizing the reference 3D shape of the input clipart and rendering it from the desired viewpoint.
The critical challenge of constructing this visual scaffold is to generate a reference 3D shape that matches the user's expectations in terms of object sizing and positioning while preserving the geometric style of the input clipart.
To address this challenge, we propose a user-assisted curve extrusion method to obtain the reference 3D shape.
We render the synthesized reference 3D shape with a consistent style into the visual scaffold. 
By following the generated visual scaffold, the users can efficiently design clipart with their desired viewpoints.
The user study conducted by an intuitive user interface and our generated visual scaffold suggests that our system is especially useful for estimating the ratio and scale between object parts and can save on average 57\% of drawing time.
\keywords{vector graphics, clipart, multi-view, hci}

\ccsdesc[500]{Computing methodologies~Parametric curve and surface models}
\ccsdesc[100]{Computing methodologies~Image processing}
\ccsdesc[300]{Human-centered computing~Graphical user interfaces}

\printccsdesc   
\end{abstract}  

%% file: introduction.tex
\section{Introduction}
Vector clipart is widely used in graphic design for compactly expressing concepts or illustrating objects in daily life. 
For example, designers communicating their design ideas with others, or presenters illustrating concepts with cliparts.
Whenever we need an icon in our slides or videos, we usually try to search it in online clipart repositories using our target keywords, such as a chair, a lamp, or an airplane. 
A common situation we encounter is that even if we are lucky enough to find clipart with satisfactory shapes and colors, sometimes it is designed from an undesired viewpoint.
In the end, we often compromise with either a satisfactory appearance or a viewpoint.

When facing the above situation, one might consider manipulating the existing clipart to the desired viewpoint via editing tools. 
However, it is challenging to draw other views of an object with an only one available view due to the individual differences of their mental rotation ability~\cite{vandenberg1978mental}.
We conducted interviews with two professional artists, and they tackle this task by first creating a rough 3D model based on the available view.
The rough 3D model can then be used as the mental prior for designing clipart from other viewpoints.
Unfortunately, this procedure requires professional training in 3D modeling and a good mental prior of the 3D object, which is almost impossible for general users.

In this paper, we propose a multi-view clipart assistive system to show unseen viewpoints of the clipart to support mental rotation.
Our system follows the drawing-by-observation technique. 
Given the input clipart, our system's goal is to automatically synthesize other unseen viewpoints of the clipart object without changing the character of the clipart.
Users can efficiently design high-quality clipart with their desired viewpoint by simply following the visual scaffold.
Our system design simulates the creation process of professional artists. 
Instead of directly synthesizing unseen views from the 2D input, we first infer a reference 3D model based on the input clipart's viewpoint.
We generate the reference 3D model with user-provided structural annotations that preserve the style of the input clipart. 
The visual scaffold is then generated by rendering the reference 3D model from the desired viewpoint. 
Finally, we also provide an intuitive user interface to assist users in their clipart creation.

\begin{figure}[t!]
\centering
\includegraphics[width=\linewidth]{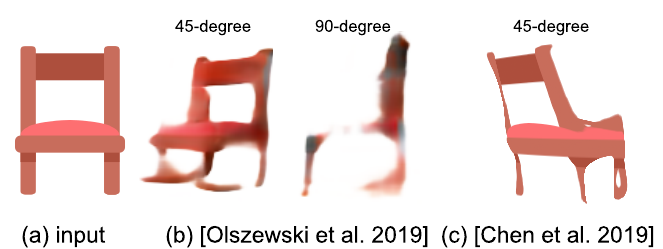}
\caption{
\cgfhl{
Given the input clipat, both (b) \mbox{Olszewski~\etal~\cite{olszewski2019transformable}} (TBN) and (c) \mbox{Chen~\etal~\cite{chen2019monocular}} fail to synthesize novel view of the input clipart. 
We use the pre-trained model for both methods provided on their webpages.}
}
\label{fig:moti_fig}
\end{figure}

The major contribution of our work is an algorithm for generating a style-consistent 3D model from the single-view clipart.
Even though there are existing works focusing on novel view synthesis~\cite{zhou2016view,sun2018multiview,park2017transformation,dosovitskiy2016learning,MV3SI}, there are two reasons that prevent us from directly applying these methods to serve our purpose.
First, these methods require substantial training datasets to learn the categorical shape priors, but there is no existing huge clipart dataset that can facilitate such a learning process.
\cgfhl{
Second, directly infer the novel view or 3D representation of the clipart object using the pre-trained models of the methods mentioned above usually lead to failure cases (\mbox{\figname~\ref{fig:moti_fig}}).
}
The reason is that the object boundaries in clipart are mainly composed of low-degree curve types such as line and arc, not the cases in existing datasets.
Due to the very different geometric styles, the predicted 3D geometry generated by previous methods would not be style-consistent with the input clipart.  
Our proposed user-assisted curve extrusion method addresses this problem by combining state-of-the-art 3D reconstruction and user-provided structural annotations to guide the curve extrusion method.
The generated 3D shape is regularized and style-consistent with the style of the input clipart.
\cgfhl{
Noted that we did not aim for proposing a novel deep learning architecture that handles vector clipart directly.
Instead, we leverage an existing 3D reconstruction method using a raster image and combine it with structural annotation to generate a regular 3D shape.
}

We conducted a user study with general users who did not have much drawing training or experience.
The purpose is to evaluate how helpful our generated visual scaffold is for designing clipart from an unseen viewpoint.
To conduct this user study, we designed a complete system for assisting general users in multi-view clipart creation.
In our intuitive user interface, we show two display areas side-by-side: the \textit{input reference area} and the \textit{drawing area}. 
In the input reference area, we show the input clipart to the user, as most users will keep referring to the input viewpoint while drawing the novel viewpoint. 
In the drawing area, the user can use both curve tools and shape tools to design the novel view clipart. 
The visual scaffold we synthesized lies under the drawing area.
With our system, most of the users could design better clipart from a novel viewpoint with a regularized visual scaffold in a small fraction of the original time.
Furthermore, most participants agree that our visual scaffold significantly reduces their cognitive workload when designing clipart from a new viewpoint.

%% file: relatedwork.tex
\section{Related Work}
\subsection{Novel-view synthesis}
To synthesize photo-realistic novel view images, most traditional methods~\cite{Soft3DReconstruction,seitz2006comparison,Kopf2013,debevec1996modeling} take multi-view images as input and infer the 3D representations explicitly to facilitate novel view rendering.
On the other hand, recent learning-based methods~\cite{zhou2016view,sun2018multiview,park2017transformation,dosovitskiy2016learning,MV3SI} provide the ability to render novel-view images using only one single image.
Some methods~\cite{dosovitskiy2016learning,MV3SI} directly synthesize pixels of the desired viewpoint, while the other methods~\cite{zhou2016view, park2017transformation,sun2018multiview} instead estimate the flow map from the input view to the desired viewpoint.
Some methods~\cite{karras2019style,nguyen2019hologan} disentangled the image factors such as viewpoint, shape, and appearance from the source image and used these factors for facilitating view synthesis.
There are also learning-based methods focusing on generating better data representation for synthesizing novel views from a single image, including  voxel and volume-based representation~\cite{drcTulsiani17,choy20163d,niemeyer2019differentiable}, point cloud~\cite{fan2017point,mo2020pt2pc}, mesh~\cite{groueix2018, Dai_2019_CVPR,liu2019soft}, depth and/or normal maps~\cite{lun20173d,li2018robust}, and parametric surface~\cite{DeepSketch}.

The goal of our system is similar to novel view synthesis from a single-view image.
However, unlike previous methods that focus on photo-realistic raster images and have many training data available, there is no existing clipart dataset (neither raster-based nor vector-based) that can facilitate such a learning-based approach.
\cgfhl{
Meanwhile, \mbox{Lopes~\etal~\cite{Lopes_2019_ICCV}} proposed a generative model for font synthesis with a deep learning framework handles vector format.
However, we can not apply this method directly for our application because their model did not handle (i) the color, (ii) shape category, and (iii) the viewpoint information.
Hence, we combine a learning-based approach on raster image with user-provided structural annotation and provide an intuitive interface to aid the creative process.
}

\begin{figure*}[t!]
\centering
\includegraphics[width=\linewidth]{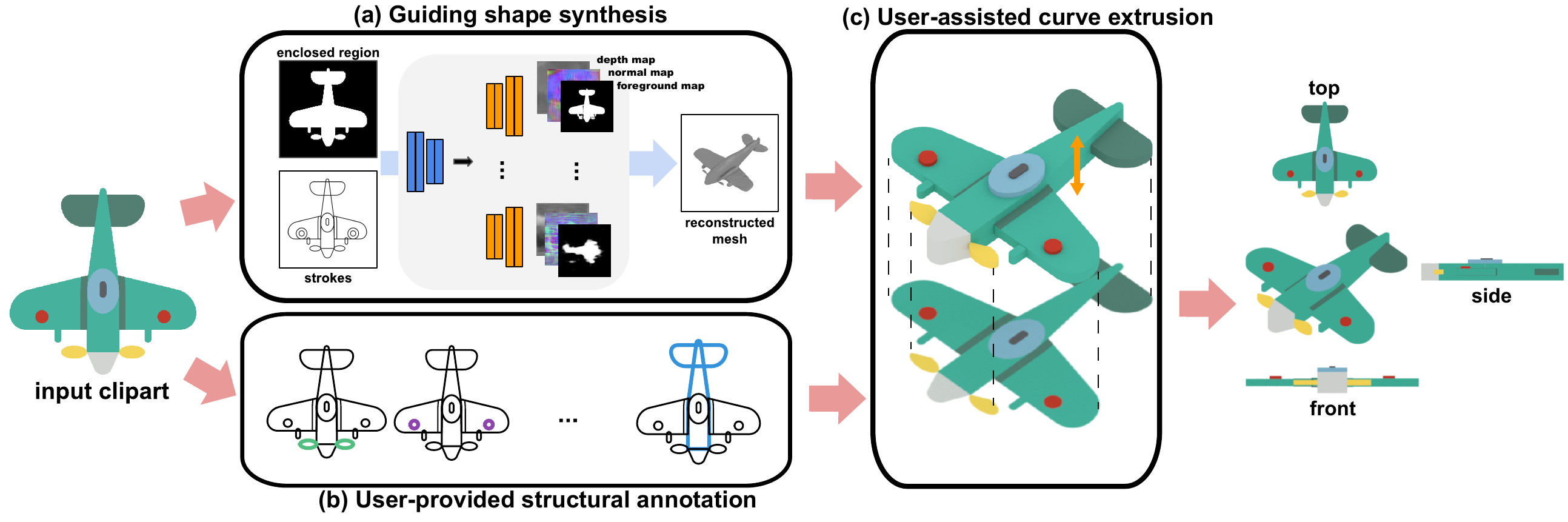}
\caption{
\cgfhl{
Given the input clipart, our method (a) use a single-view 3D reconstruction method to synthesize a guiding shape.
(b) The user can annotate the input clipart to provide the structural information. 
(c) Our method extrude the curves in the input clipart adaptively by leveraging the predicted guiding shape and the user-provided structural annotations.
}
}
\label{fig:overall_flow}
\end{figure*}

\subsection{Clipart synthesis}
Vector clipart can be synthesized by vectorizing existing raster images or designing from scratch.
Commercial products~\cite{illustrator, vector_magic} provide robust image vectorization that simultaneously tackles both image segmentation and curve (segment boundary) fitting problem.
However, the clipart style usually differs from the natural image (\eg~flat shadings and rounded shapes), which makes the vectorized natural image not suitable for synthesizing clipart.
Some previous works have focused on vectorizing pixel arts~\cite{hqx,kopf2011depixelizing,hoshyari2018perception}.
Kopf~and~Lischinski~\cite{kopf2011depixelizing} proposed a dedicated method that focuses on resolving the topological ambiguities for pixel arts vectorization.
Hoshyari~\etal~\cite{hoshyari2018perception} used human perceptual cues to generate boundary vectorization that can better match viewers' expectations.
Liu~\etal~\cite{Liu:2016:DI} designed an interactive system to synthesize novel clipart by remixing existing clipart in a large repository.

\subsection{Assisting authoring tools}
Content authoring is a fundamental problem in Computer Graphics.
Many previous works focus on assisting the authoring workflow. 
Among them, many works utilize the personal editing histories to assist 2D sketch~\cite{Xing:2014:APR}, 3D shape sculpturing~\cite{Peng:2018:AS}, and viewpoint selection~\cite{Chen:2014:HAV}.
Besides the history-related assists, Xing~\etal~\shortcite{Xing:2016:EIT} utilizes the energy strokes from professional artists to facilitate the authoring process of 2D animation.
Lee~\etal~\shortcite{Lee:2011:SRU} proposed a guidance system that dynamically updates a shadow image underlying the user's strokes while the user is drawing.
Hennessey~\etal~\shortcite{how2sketch} automatically generates a step-by-step tutorial for drawing 3D models.
Ryan~\etal~\shortcite{SIJSW07,SKSK09} provide construction lines and primitives as visual scaffolds for better drawing 3D objects.
Our system shares the same spirit and provides a visual scaffold to support users for designing clipart.
The major difference is that we provide the accurate rendered image as a visual scaffold instead of hints such step-by-step as construction lines.
According to the feedback from our study, the visual scaffold of the final rendered image can provide an overview of the shape in the beginning so that they can better arrange the order of drawing.

\subsection{Geometric stylization}
Creating 3D shapes with different styles of geometric has drawn a lot of attentions including Japanese manga style~\cite{Shen:2012:SDM}, Lego brick style~\cite{Luo:2015:LOL}, abstraction style~\cite{mzlsgm_abstraction_siga_09,Yumer:2012:COA}, and cubic style~\cite{liu2019cubic}.
Lun~\etal~\shortcite{Lun:2015:StyleSimilarity,Lun:2016:StyleTransfer} analyze the style similarities between shapes, and transfer different styles between shapes.
Unlike the previous works that manipulate the input 3D shapes using mesh deformation techniques, our goal is to synthesize the 3D shape from the input clipart.
We combine the single-view 3D mesh reconstruction method and user's structural annotations to guide the curve extrusion process and synthesize reference 3D shape that matches the input clipart in both geometric and appearance styles.

%% file: method.tex
\section{Visual Scaffold Synthesis}
Our system includes two major components: (i) visual scaffold synthesis, which synthesizes the input clipart under a desired viewpoint, and (ii) the drawing user interface, which displays the synthesized visual scaffold beneath the user’s drawing canvas to aid the drawing process.
\input{singleview}
\input{user-assisted}

%% file: singleview.tex
Given the input clipart consists of multiple closed paths $S=\{C_0, C_1,.. C_{n-1}\}$ under input viewpoint $\theta^{i}$, the goal of this step is to synthesize the visual scaffold that aids the users in designing the clipart from an unseen viewpoint $\theta^{u}$ efficiently.
To achieve this goal, we design our method by addressing the following three aspects:
\begin{enumerate}
\item the shape in the visual scaffold image has to match the user's imagination for the input clipart from viewpoint $\theta^{u}$.
\item the appearance style of the visual scaffold has to match the input clipart.
\item the geometric style of the shape in the visual scaffold has to match the input clipart.
\end{enumerate}
If the shape in the visual scaffold conflicts with the user's imagination (\ie~violate (1)), it will hinder the creative process instead of providing useful aids.
In the meanwhile, if the geometric and appearance style did not match the input clipart (\ie~violate (2) and (3)), the clipart created by the user will not be depicted as clipart.
To address the above observations, we propose a user-assisted curve extrusion method to generate a reference 3D shape from the input clipart.
The user is allowed to provide structural annotations on the input clipart to indicate the 3D structure information.
As illustrated in \figname~\ref{fig:overall_flow}, our method generates a reference 3D model $M$ by following the guidance of both user's structural annotations and the guiding mesh $M^{G}$, which is reconstructed by using a single-view mesh reconstruction method based on the input clipart $S$.
We search for optimal thickness and transformations of extrusion to simultaneously match users' structural annotations, while best interpreting the guiding shape $M^{G}$.
\cgfhl{
Noted that we use the rasterized curves of the input clipart $S$ for inferring the guiding mesh.
The main reason for choosing the raster-based method is because unlike the mature architecture for raster image, such as CNN (convolutional neural network), there is no mature learning-based method that handles the vector clipart well.
}
\minorhl{
Moreover, this paper aims not to propose a novel deep learning architecture that directly handles vector clipart. 
Even though there exists relevant work~\mbox{\cite{lopes2019learned}} that handles vector format data using a deep generative model, however, their model did not consider (i) the color, (ii) shape category, and (iii) the viewpoint information. These pieces of information are vital to our application. So we choose to leverage more mature methods proposed on raster image and focus on our application of aiding users to design clipart under unseen viewpoint.
}

\subsection{Single-view guiding shape synthesis}
\label{sec:singleview}
Performing 3D reconstruction from single-view input has been a very challenging task.
In recent years, with the fast advances of the deep learning methods, the state-of-the-art method has shown the possibility of learning a descriptive latent representation for category-specific objects and synthesizing the corresponding shape represented by voxels, point clouds, or meshes.
In this work, we reconstruct the guiding 3D shape $M^{G}$ based on the input clipart $S$ by following Lun~\etal~\shortcite{lun20173d}.
\begin{figure}[t!]
\centering
\includegraphics[width=0.8\linewidth]{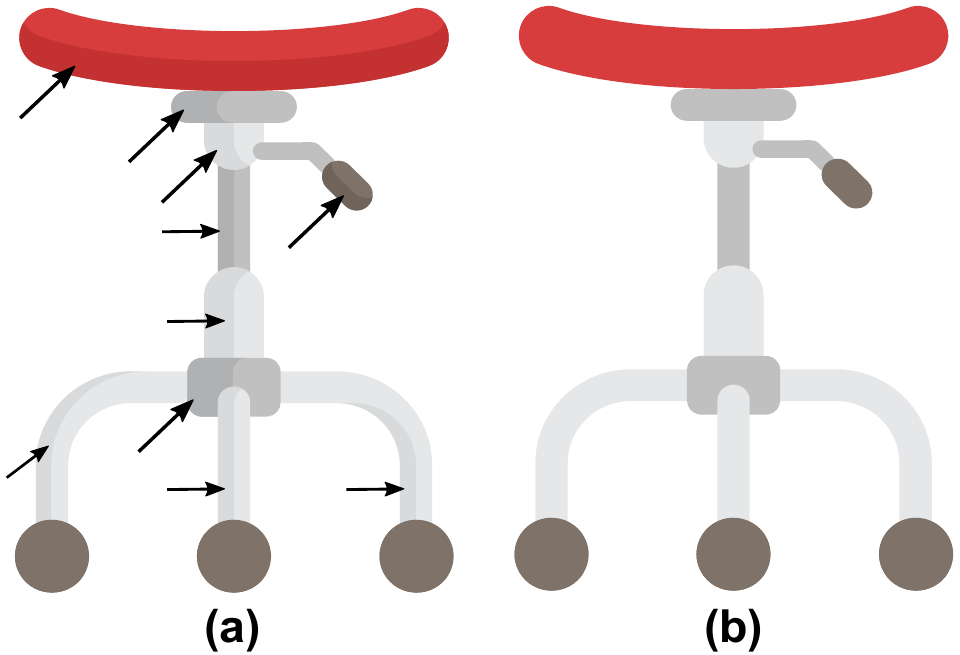}
\caption{
(a) Some closed paths (pointed by arrows) in the clipart represent shading instead of geometry.
(b) After removing all the shading path, the remaining closed paths represent geometry only.
}
\label{fig:shading_path}
\end{figure}
In their method, they use an encoder-decoder architecture where there are multiple (12 in their paper) decoders that predict geometric data, including (i) the normal maps, (ii) the depth maps, and (iii) foreground probability maps for each viewpoint.
For each input sketch, they applied Poisson surface reconstruction~\cite{Kaz06Poisson} on the predicted geometric data to generate the mesh.
The entire pipeline is illustrated in \figname~\ref{fig:overall_flow}(a).

We observe several differences between our input clipart and the original sketch data used in \cite{lun20173d}, which prevents us from directly applying their method to our input data.
Thus we perform the following preprocesses before we feed the input clipart into their method: %proposed in \cite{lun20173d}:
\begin{description}[style=unboxed,leftmargin=0cm]
\item[Shading path removal] 
One characteristic of the clipart is that it usually describes both \textit{appearance} and \textit{geometry} in the same file.
Unlike the 3D model file, which usually describes the geometry only, its appearances are defined through different material or texture files.
However, the path elements in the clipart sometimes represent the shading, \eg~reflection (see \figname~\ref{fig:shading_path})), which is not directly representing the shape.
In this work, we focus on aiding the users to design the geometry part of the clipart, so we categorized path elements in the clipart into two categories, \ie~geometry path and shading path. 
Furthermore, we remove the shading path before we perform the 3D reconstruction.
\cgfhl{
Currently, this process is performed manually.
The main reason is that we do not have lighting information; thus, we can't distinguish path type only based on its curve geometry and color information.
}
\item[Color removal]
We remove the filled color of each closed path in the input clipart.
The remaining curve outlines are served as the sketches' input to the method proposed in \cite{lun20173d}.
\end{description}

\begin{figure}[t!]
\centering
\includegraphics[width=1.0\linewidth]{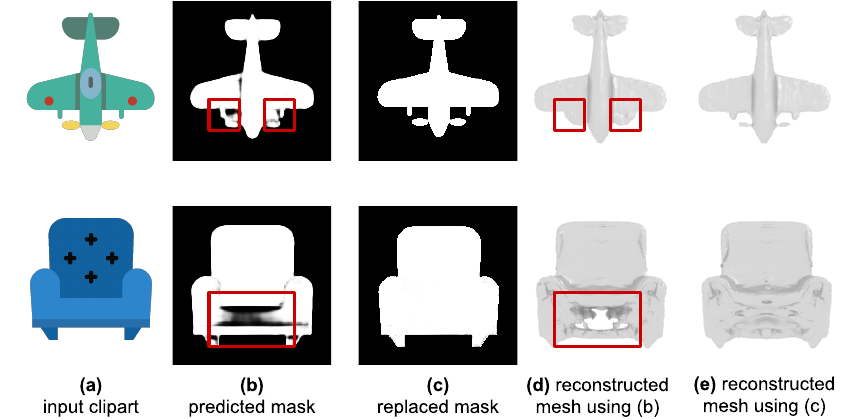}
\caption{By replacing the predicted noisy foreground probability mask (b) with the known mask derived from the input clipart (c), we can greatly improve the quality of the reconstructed meshes.
}
\label{fig:mask_result}
\end{figure}

After the preprocessing, we feed the processed curve outlines into \cite{lun20173d} and obtained the predicted geometric data.
Due to the large differences of geometric styles between clipart and the shapes in their training dataset, the resulted point cloud constructed by the predicted geometric data is usually broken, as shown in \figname~\ref{fig:mask_result}(b).

One possible solution is to re-train their model with a huge clipart dataset; however, collecting large numbers of clipart with multiple viewpoints is expensive and time-consuming.
As a result, we choose to use the pre-trained model provided by \cite{lun20173d} as our inference model.
Thus, the quality of predicted shapes is usually worse and noisier than the predicted results using the 3D shape sketches that capture the real-world shape style.
To further filter out noisy predicted shape data, we use the input clipart to create an auxiliary foreground mask (\figname~\ref{fig:mask_result}(c)) from the viewpoint of the input clipart.
As we observed, this step turns out pretty helpful for compensating for the bad quality due to the geometric style difference. 
Please see \figname~\ref{fig:mask_result} for reconstructed shapes with and without this auxiliary foreground mask.

%% file: user-assisted.tex
\subsection{User-assisted curve extrusion}
\begin{figure}[t!]
\centering
\includegraphics[width=0.9\linewidth]{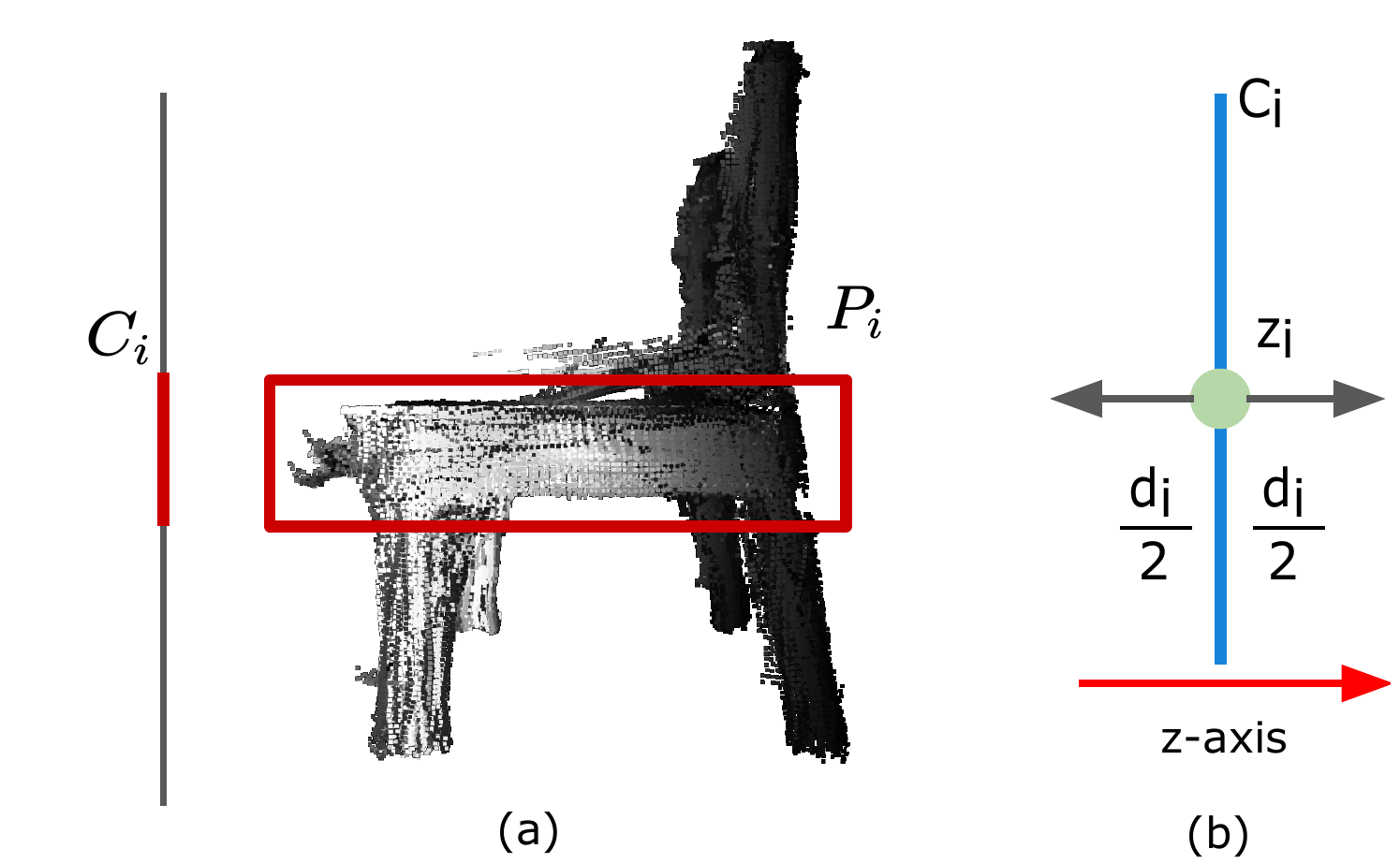}
\caption{
(a) For each curve $C_i$ (the red segment on the left), we project vertices on the guiding shape $M^{G}$ and obtain a pointset $P_i$ that is enclosed by $C_i$ (inside the red rectangle).
(b) Our extrusion parameterization.
}
\label{fig:opt_fig}
\end{figure}
After obtaining the guiding shape $M^{G}$, we want to leverage it to guide the curve extrusion process to make a 2D curve into a 3D volume.
Curve extrusion is widely used in designing the CAD model where the designers usually design layout the shape in 2D and extrude certain thickness along the direction into solid volumes.
Our goal is to extrude all the closed paths in the input clipart $S$ into volumes and transform them to cover as many vertices in the guiding shape $M^{G}$ as possible.
We assume the extrusion axis is along the z-axis (\ie~the input clipart lying on the xy-plane).
And we parameterize the extruded volume $V_i$ of a closed path $C_i$ using the following parameter: (i) the extrude thickness $d_i$, and (ii) the z-coordinate $z_i$ of the centroid of the extruded volume (see \figname~\ref{fig:opt_fig}(b) for illustration).

\begin{figure}[t!]
\centering
\includegraphics[width=\linewidth]{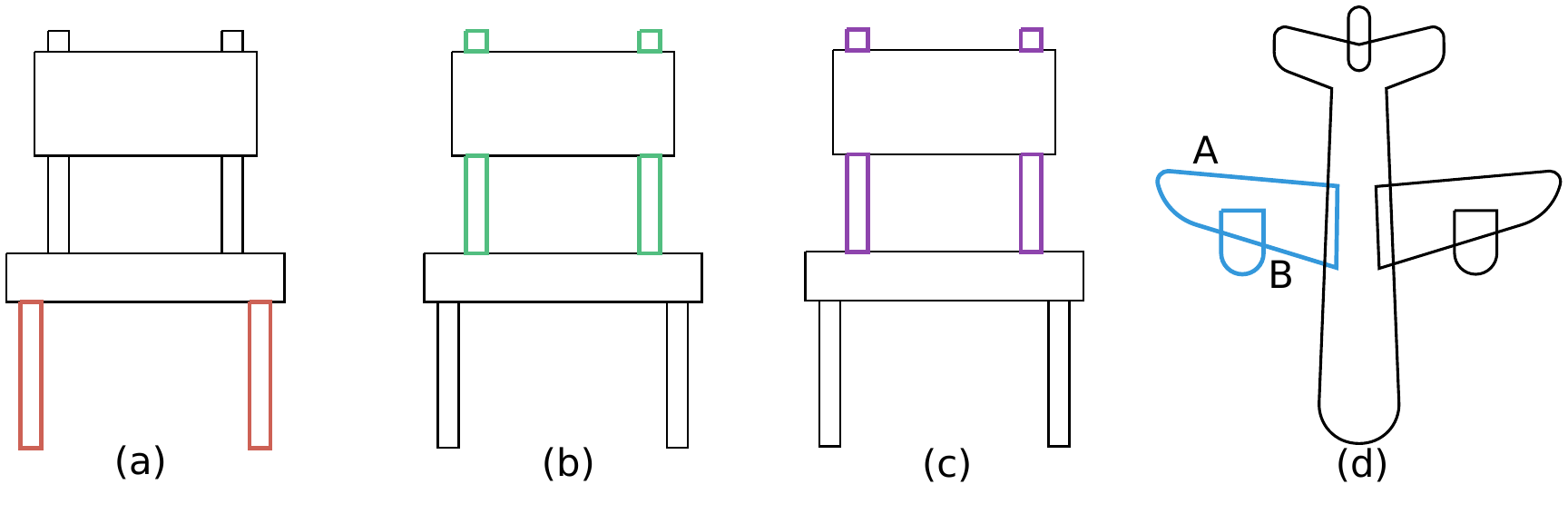}
\caption{The user can annotate four different types of structural information. 
(a) The paths of chair legs (in red) are annotated as \textit{multiple objects}. 
(b) Four paths (in green) are annotated as \textit{same thickness}. 
(c) Four paths (in purple) are annotated as \textit{same depth}.
(d) Path A is annotated as it is in front of path B (blue).
}
\label{fig:annotation}
\end{figure}

\begin{description}[style=unboxed,leftmargin=0cm]
\item[User-assisted structural annotation] 
We allow the users to annotate each closed path in the input clipart with some structural properties.
These annotations enable the users to assign the 3D information they depicted.
We provide the following four types of annotations:
\begin{itemize}
    \item \textbf{multiple objects}: for each closed path $C_i$ in 2D, there might be multiple objects in 3D which are occluded. 
    We provide this annotation to the user to assign the number of objects $N$ they think exists. 
    We duplicate the annotated closed path $N$ times before doing the extrusion.
    \item \textbf{same thickness}: for each pair of closed paths $C_i$ and $C_j$, the user can enforce them to have the same extrude thickness (\ie~$d_i = d_j$). 
    \item \textbf{same depth}: for each pair of closed paths $C_i$ and $C_j$, the user can enforce the z-coordinate of the centroid ($z_i$ and $z_j$) of the extruded volume to be the same.
    \item \textbf{depth order}: by default, we will leverage the input clipart's layering as the depth orders between closed paths. 
    However, the user can assign the desired depth order to overwrite the default ordering.
\end{itemize}
\end{description}
Please see \figname~\ref{fig:annotation} for illustrations of each structural annotations.

For each closed path $C_i$, we project the vertices of the guiding shape $M^{G}$ back onto the xy-plane and obtain a pointset $P_i$ that is enclosed by $C_i$ (see \figname~\ref{fig:opt_fig}(a)).
\minorhl{
For the closed path annotated as multiple objects, we perform a clustering operation on the enclosed pointset $P_i$ based on the 3D points' positions, \eg~if a closed path is duplicated twice, we generate two pointsets $P_i^0$ and $P_i^1$.
And we assign each pointset to each duplicated closed path.
}
\cgfhl{
And we obtain the best extruded volumes $\mathbf{V}=\{V_1, ..., V_n\}$ by optimizing the following geometry approximation cost function:
}
\begin{align}
\text{minimize}& \quad E^{cover} + \omega E^{thickness}\label{eq:all_cost}\\
\text{subject to }
&\quad z_i-z_j = 0 \qquad \textbf{same depth} \label{eq:depth_cost}\\
&\quad z_i-z_j > 0 \qquad \textbf{depth order} \label{eq:order_cost}\\
&\quad  d_i-d_j = 0 \qquad \textbf{same thickness} \label{eq:thick_cost}
\end{align}
We define the volume coverage cost ($E^{cover}$) as:
\begin{align}
dist(x, V_i) &= \begin{cases}
0, \; \text{if }x \; \text{is inside} \; V_i\\
\min_{q\in \Omega(V_i)} \|x-q\|, \; \text{otherwise},
\end{cases} \\
E^{cover}_i &= \sum_{x\in P_i} dist(x, V_i) \\
E^{cover} &= \sum_{i=1}^{n} E^{cover}_i
\label{eq:cover}
\end{align}
where $x$ is a point belongs to $P_i$, $n$ is the number of closed paths in input clipart $S$, and $\Omega(V_i)$ represents the surface of the extruded volume.
And we define the thickness cost ($E^{thickness}$) as :
\begin{align}
E^{thickness} &= \sum_{i=1}^{n} \|d_i\|^2.
\end{align}

The intuition of optimizing \eqname~\ref{eq:all_cost} is to encourage the volume to cover as many enclosed points as possible and keep the extrude thickness small.
While minimizing the cost function (\eqname~\ref{eq:all_cost}), the structural annotation is formulated into constraints (\eqname~\ref{eq:depth_cost} to \ref{eq:thick_cost}) that enforce the structural relationship between closed paths in the input clipart $S$.
\cgfhl{
The \mbox{$\omega$} we used in \mbox{\eqname~\ref{eq:thick_cost}} is set as $100$ by doing a grid search over a set of possible values on two verification clipart. 
We found out the value better balance the scale between both terms and discourage the over-thickness extrusion results.
}

\begin{figure}[t!]
\centering
\includegraphics[width=\linewidth]{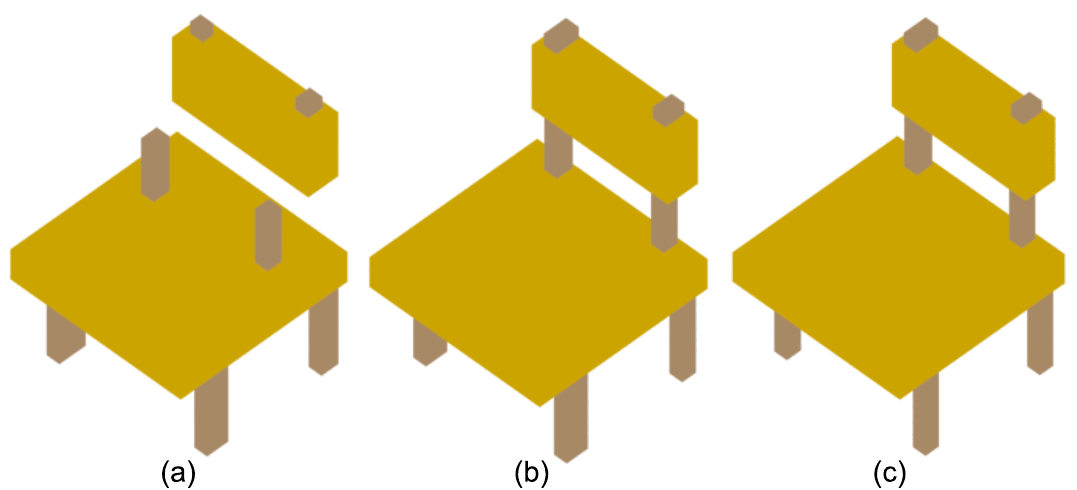}
\caption{
The result after (a) initialization of the thickness ($d$) and the depth value ($z$), (b) resolving depth value, and (c) resolving thickness.
}
\label{fig:opt_process}
\end{figure}

\begin{description}[style=unboxed,leftmargin=0cm]
\item[Optimization method] 
To simplify the optimization problem, we propose to use a greedy method to obtain the optimized result.
To optimize \eqname~\ref{eq:all_cost} for each closed path $C_i$, we first fit an oriented bounding box (OBB) $O_i$ for the enclosed pointset $P_i$.
For each $d_i$, we limit the possible values as three side lengths of $O_i$. 
We initialize $d_i$ as one of the side lengths with the minimum $E^{cover}$ value.
And we initialize $z_i$ as the centroid of $P_i$.

Next, we resolve the structural constraints (from \eqname~\ref{eq:depth_cost} to \ref{eq:thick_cost}) one by one.
For \textbf{same depth} constraint, for closed paths $C_i$ and $C_j$ that are annotated as same depth, we move one of their extruded volume (\ie~$V_i$ or $V_j$) that leads to the smaller $E^{cover}$.
For \textbf{depth order} constraint, we also choose one of the extruded volumes between $V_i$ and $V_j$ and move the z coordinate of the chosen volume (either $z_i$ or $z_j$) that leads to smaller $E^{cover}$. 
Finally, for \textbf{same thickness} constraint, we also choose one of the extruded volumes between $V_i$ and $V_j$ and set their thickness value of the chosen volume (either $z_i$ or $z_j$) that leads to smaller $E^{cover}$. 
In \figname~\ref{fig:opt_process}, we show an illustrative example of the optimization process.
\minorhl{
The main reason for utilizing the greedy method instead of the traditional gradient-based optimization method is that the two sets of variables (\mbox{\ie~the thickness} $d_i$ and the centroid of extruded volume $z_i$) related to each other.
Specifically, the different centroid of extruded volume leads to different optimal thickness and vice versa.
Although this kind of problem is studied and solvers are proposed for different applications~\mbox{\cite{liu2008local,bouaziz2012shape}}, their convergence is not guaranteed, and it is likely to extend the running time.
In our experiment, our greedy method can achieve better results in most of the cases.
}
\end{description}

%% file: userinterface.tex
\section{User Interface}
\label{sec:ui}
\begin{figure}[t!]
\centering
\includegraphics[width=\linewidth]{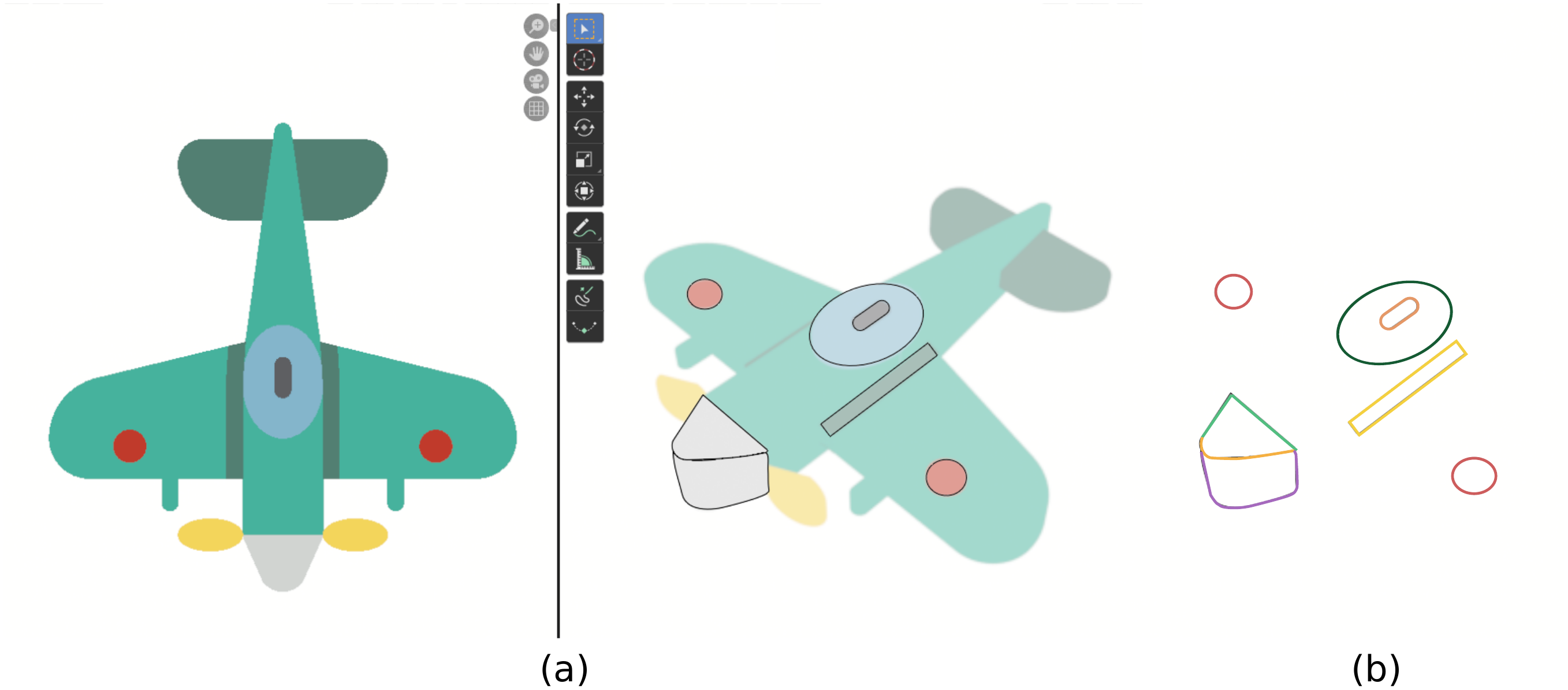}
\caption{
(a) Screenshot of our user interface.
On the \textit{reference} area (left), we show the input clipart.
On the \textit{canvas} area (right), the visual scaffold is overlaid under the canvas.
The user can use curve tools and shape tools to design clipart from the desired viewpoint.
(b) In the current design session, the user used different curves and shapes (represented in different colors), including line, arc, bezier freeform tools, rectangle, ellipse/circle, and rounded rectangle.
\cgfhl{
Please see the attached video for the detail of our user interface.
}
}
% \vspace{-2.5mm}
\label{fig:ui_screenshot}
\end{figure}

\figname~\ref{fig:ui_screenshot}(a) illustrate our user interface, which consists of two separate areas.
The \textit{reference} area shows the input clipart, which acts as the reference viewpoint for designing clipart from the desired viewpoint.
The \textit{canvas} area is the place that the user designs the clipart using curve and shape tools.
In the \textit{canvas} area, we overlay the visual scaffold under the canvas, and the user can decide if he/she wants to follow the scaffold or not.

In the \textit{canvas} area, we provide two sets of drawing tools, \ie~curve tools and shape tools.
In the curve tools, we provide (i) line, (ii) arc, and (iii) freeform tools.
The reason for choosing line and arc is because we observed that many clipart could be described solely using lines and arcs.
In the shape tools, we provide (i) rectangle, (ii) ellipse/circle, and (iii) rounded rectangle.
These shape primitives are chosen because they are often used to design clipart from canonical viewpoints.
Please see \figname~\ref{fig:ui_screenshot}(b) for the illustration of different tools.

In the \textit{canvas} area, the user can first sketch out the outline of the shape he/she wants to draw.
The sketch can be hidden during the designing process.
The user can also draw separate curves and primitives in different layers and move layers forward and backward if he/she wants to.

%% file: result.tex
\section{Results and Evaluation}
We used our user-assisted curve extrusion method to generate several man-made 3D shapes of different clipart.
For each input clipart, we compared the 3D shape was generated by our method with the following methods:
\begin{description}[style=unboxed,leftmargin=0cm]
\item[Single-view sketch shape reconstruction]
The guiding shape we used in our method was generated by \cite{lun20173d}.
\item[Sketch-based shape retrieval]
We perform a sketch-based shape retrieval using input clipart.
The goal is to retrieve the most similar shape in ShapeNet~\cite{chang2015shapenet} dataset.
For each shape in the ShapeNet dataset, we extracted their outlines using Suggestive Contours~\cite{DeCarlo:2003:SCF} as candidate images.
And we extract the strokes of each input clipart as query and compute the similarities between the clipart strokes and contour of each shape using Shape Context descriptor~\cite{belongie2001shape}.
\item[Artist creation] 
We asked an artist to design the clipart from unseen viewpoints given the input clipart, and he generates a reference 3D shape as his reference for each clipart.
\end{description}
We assigned colors for both results using single-view sketch shape reconstruction and shape retrieval from the input clipart.
We perform shape registration between the rendered contour image to the input clipart to establish the correspondences.
And we project the color obtained by correspondences onto the retrieved 3D shape.
\minorhl{
For propagating the colors onto all triangle \mbox{$T$} of the retrieved 3D mesh, we use a formulation of a Markov random field (MRF) problem as follow:
}
\begin{align}
E(f) = w_{data} \sum_{t\in T} \mathcal{D}(t, f_t) + w_{smoothness} \sum_{t, s \in \mathcal{N}} \mathcal{S}_{t,s}(t,s,f_t, f_s),
\label{eq:Ef}
\end{align}
\minorhl{
where $f$ is the function assigns the color label to each triangle $t\in T$, $f_t$ and $f_s$ are color labels assigned to triangle $t$ and $s$, and $\mathcal{N}$ is the set of all pairs of triangles sharing edges.
And we define the data term (\mbox{\ie~$\mathcal{D}$}) as follow:
}
\begin{align}
\mathcal{D}(t, f_t)=
\begin{cases}
\|c(t)-\bar{c}(t)\|^2_2, \quad \text{if color } \bar{c}(t) \text{ is assigned to triangle } t\\
0, \quad \text{otherwise}\\
\end{cases}
\end{align}
\minorhl{
And the smoothness term (\ie~$\mathcal{S}_{t,s}$) is used to measure the spatial consistency of neighboring triangles and is defined as follow:
}
\begin{align}
\mathcal{S}(t,s,f_t,f_s)=
\begin{cases}
0, \quad \text{if }f_t = f_s \\
-\log(\theta_{t,s}/\pi)\psi_{t,s}, \quad \text{otherwise},
\end{cases}
\end{align}
\minorhl{
where $\theta_{t,s}$ and $\psi_{t,s}$ are the dihedral angle and the centroid distance between $t$ and $s$, respectively.
In all the results shown in the paper, we set $w_{data}=1.0$ and $w_{smoothness}=10$.
And we solved \mbox{\eqname~\ref{eq:Ef}} using the graph cut algorithm~\mbox{\cite{boykov2004experimental}}.
Noted that the assigned colors are merely for convenient comparison.
}

\begin{figure*}[t!]
\centering
\includegraphics[width=\linewidth]{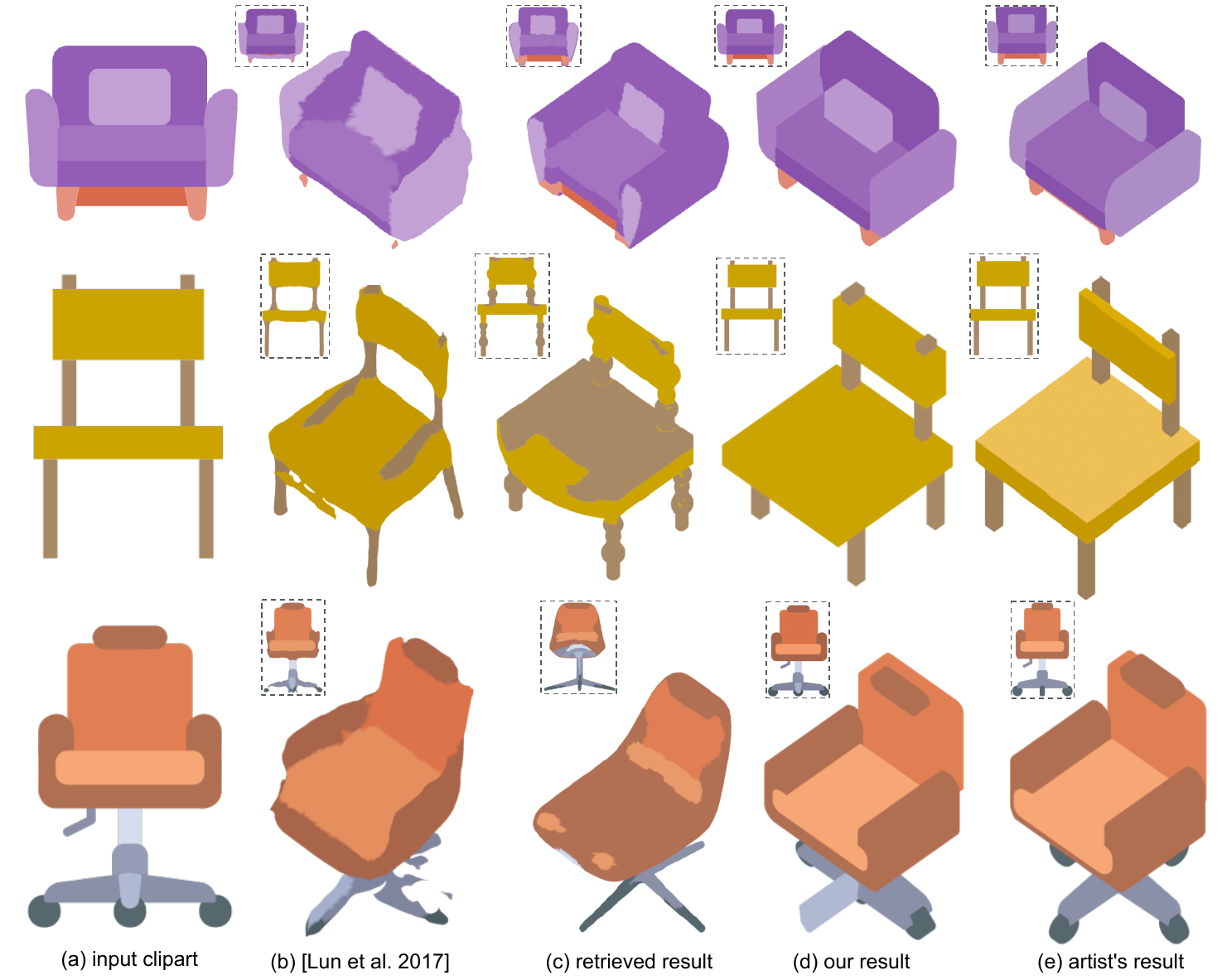}
\caption{We compared \minorhl{the reconstructed 3D chair shapes} using our method with the results generated by \cite{lun20173d}, the results using shape retrieval (see the detailed explanation in the main text), and the artist's created 3D chair shapes.
Our reconstructed 3D shapes preserve geometric and appearance styles of the input clipart compared to the results generated by other automatic methods.
The shape is more regular, and there are few incomplete structures in the reconstructed 3D shape.
In the meanwhile, our reconstructed 3D shapes are very close to the artist's created shapes.
}
\label{fig:chair_result}
\end{figure*}

\begin{figure*}[t!]
\centering
\includegraphics[width=\linewidth]{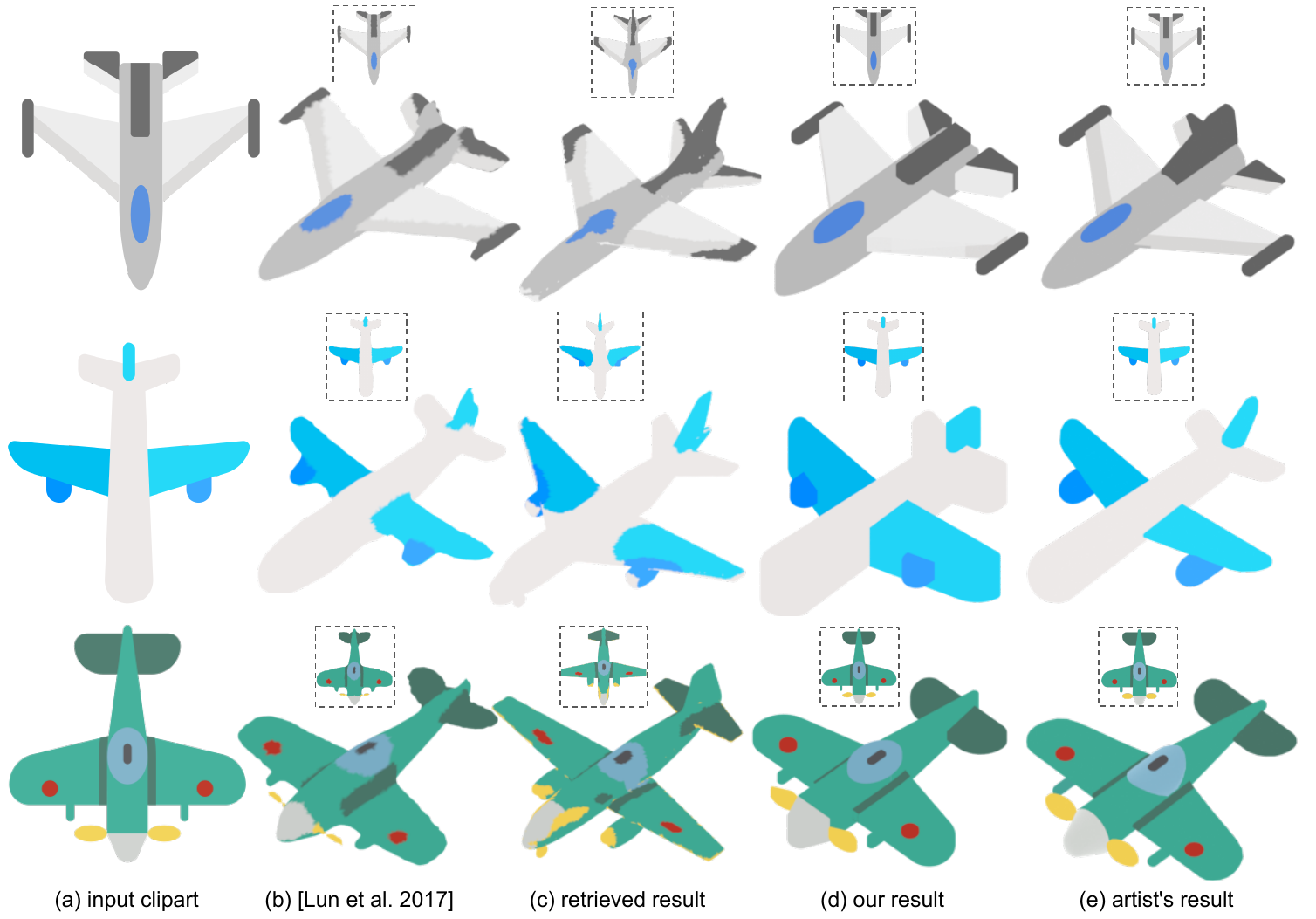}
\caption{We compared \minorhl{the reconstructed 3D airplane shapes} using our method with the results generated by \cite{lun20173d}, the results using shape retrieval (see the detailed explanation in the main text), and the artist's created 3D airplane shapes.
Our reconstructed 3D shapes preserve geometric and appearance style of the input clipart compared to the results generated by other automatic methods.
The shape is more regular, and there are few incomplete structures in the reconstructed 3D shape.
In the meanwhile, our reconstructed 3D shapes are very close to the artist's created shapes.
}
\label{fig:airplane_result}
\end{figure*}

We show all the results of compared methods in \figname~\ref{fig:chair_result} and \figname~\ref{fig:airplane_result}.
Noted that we only show 3D shapes from upper $45^{\circ}$ in the main paper, and we show the rendered image from canonical views in the supplemental material.
For all shapes, the geometric and appearance styles of our reconstructed results are more consistent with the input clipart compared to the results generated by \cite{lun20173d} and the retrieved results.
And compared to the artist's results, it took him on average 1 hour to create the 3D shapes for chair cases and 1.5 hours for airplane cases. Our method automatically produces qualitatively similar results in a fraction of this time (for chair cases, 4 mins including user annotation and extrusion optimization, and 6.5 mins for airplane cases on average).

\minorhll{
To further analyze our results, we perform a qualitative evaluation to examine 3D shapes generated by our method compared to the results of other methods.
We conduct a comparative user study by showing the input clipart, 3D shapes generated by our method and the alternative methods side-by-side. 
Then we asked the participants which one better illustrates the underlying geometry.
More specifically, we indexed the input clipart as (A) and our method and alternative as (B) or (C).
Then, the participant answers: ``Which of the two images on the bottom (B) or (C) better represent the underlying 3D structure of the input clipart (A)?''. 
All images are rendered from the upper $45^{\circ}$. 
We compared our results to three alternative methods, including \textit{sketch shape reconstruction}~\mbox{\cite{lun20173d}}, \textit{shape retrieval}, and \textit{artist creation}.
}

\begin{figure}[t!]
\centering
\includegraphics[width=\linewidth]{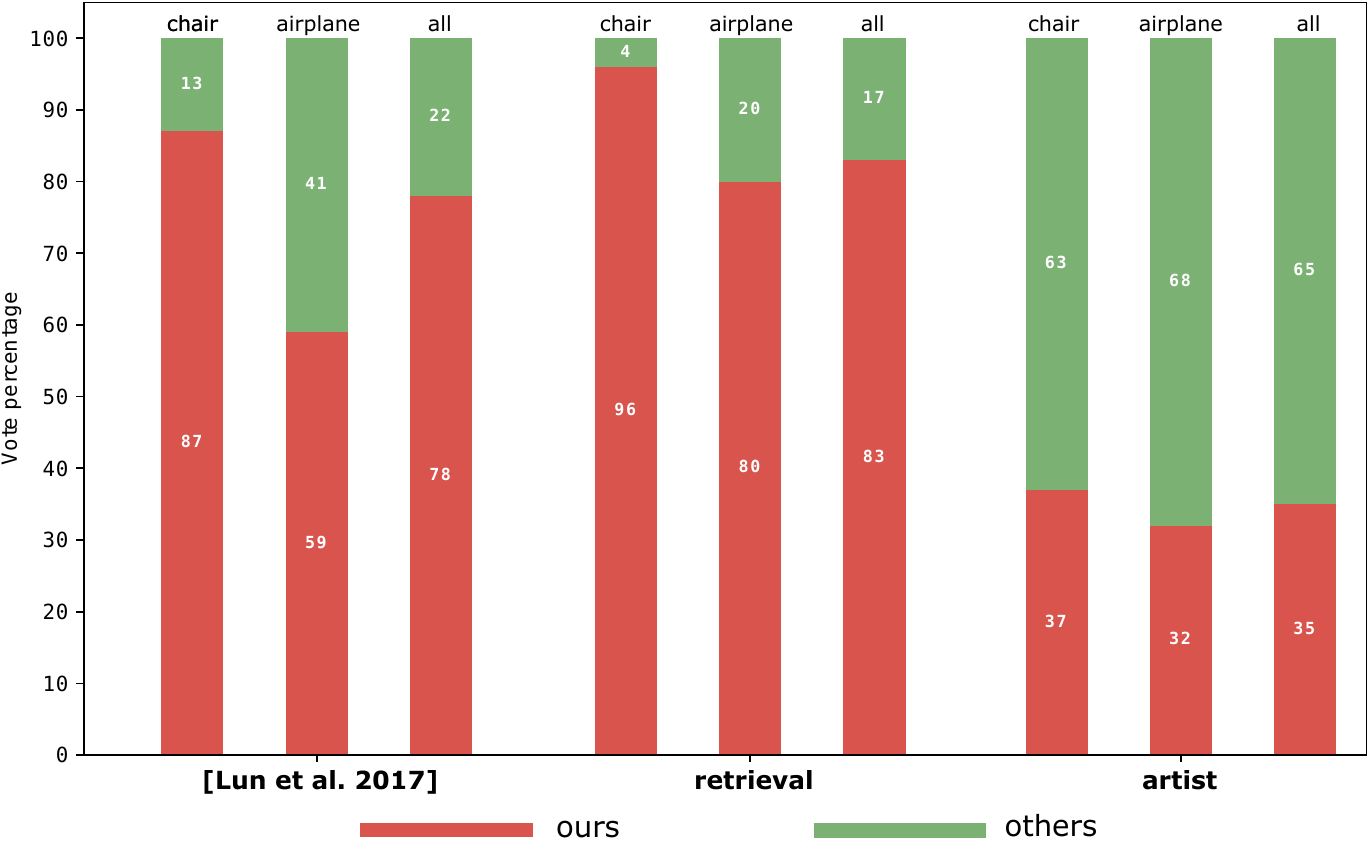}
\caption{
\minorhll{
We show the comparative statistics between our method, two alternative methods, and the artist's creation.
The result 3D shapes of our method represent the underlying 3D geometry of the input clipart significantly better than other alternative methods ($78\%$ v.s. $22\%$ and $83\%$ v.s. $17\%$).
Compared with the artist creation, there are still $35\%$ of votes in favor of our result 3D shapes, which suggests that our method can generate 3D shapes with decent quality.
}
}
\label{fig:amt_comparison}
\end{figure}

\minorhll{
We conducted this survey on Amazon Mechanical Turk (AMT), and the order of all questions was given randomly for each participant.
We excluded the response that the answer consistency is lower than $60\%$ of all questions.
In the end, 43 out of 55 participants meets our criteria. 
As shown in \mbox{\figname~\ref{fig:amt_comparison}}, $78\%$ of votes prefer our result over that of single-view sketch shape reconstruction~\mbox{\cite{lun20173d}} and $83\%$ of votes prefer our result than retrieved result.
We compare our result 3D shapes with the artist's creation.
In this case, most votes in favor of artist's creation show some room for improvement of our reconstruction method.
However, our result 3D shapes still obtain $35\%$ of votes, which suggests that our method can generate 3D shapes with decent quality.
}

\subsection{Drawing user study}
We have conducted a preliminary user study to evaluate how the visual scaffold can help users design clipart under unseen viewpoints.
\minorhll{
This study did not aim to evaluate the method of synthesizing the visual scaffold.
The participants only use the prepared visual scaffold generated by the authors.
The major reason is that to fairly evaluate how the visual scaffold assists the user in designing clipart, we have to use the same visual scaffold for different participants.
}

\begin{figure}[t!]
\centering
\includegraphics[width=\linewidth]{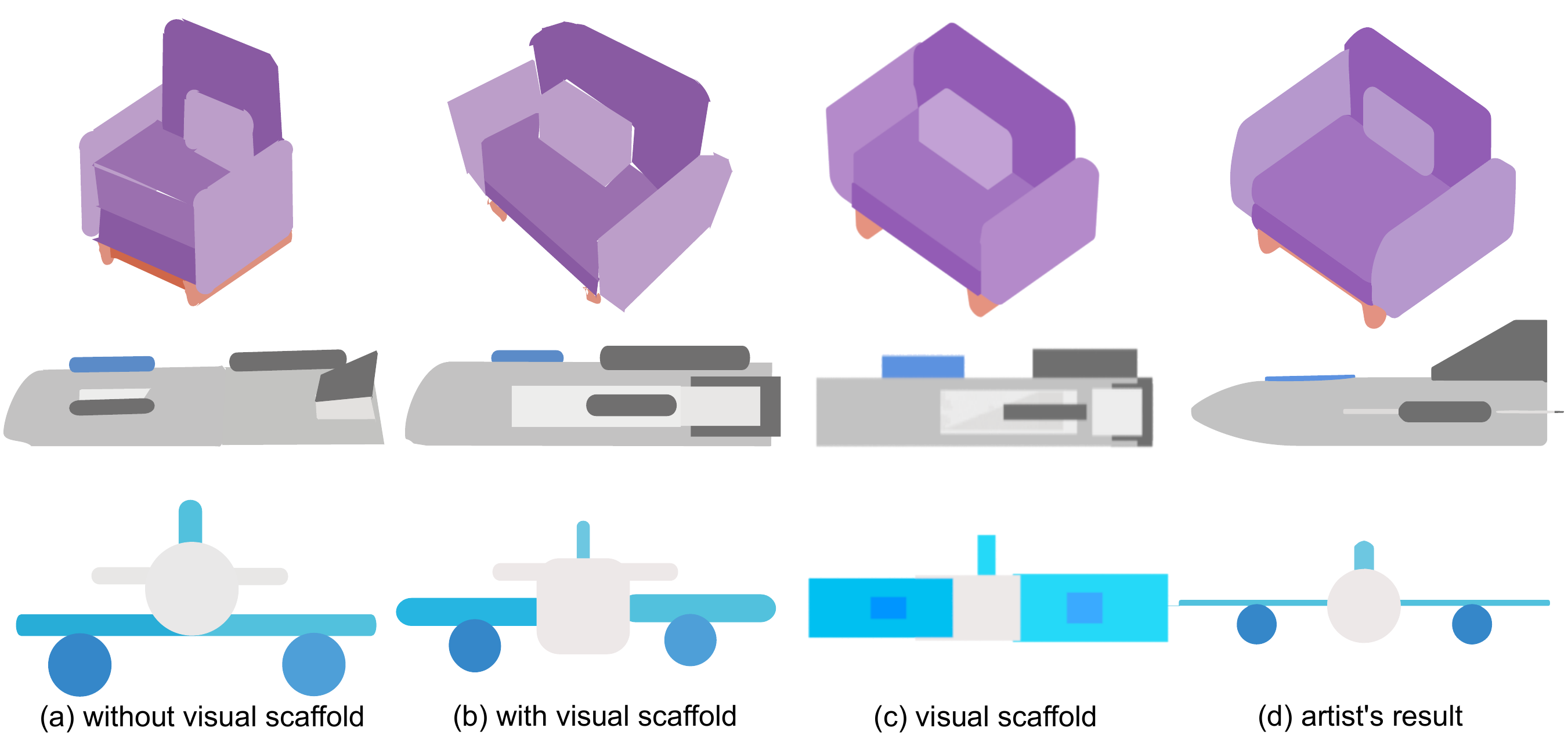}
\caption{
We compare clipart from unseen viewpoints designed by participants (a) with and (a) without our visual scaffold.
The visual scaffold is shown in (c).
With the aids of the visual scaffold, 
the participants can (i) capture the ratios between different parts of the target object from unseen viewpoints,
and (ii) place different parts in better positions with the aids of our visual scaffolds.
}
\label{fig:user_result}
\end{figure}

\begin{description}[style=unboxed,leftmargin=0cm]
\item[Setup] All tasks were conducted on a 12-inch Surface Pro laptop with a Surface stylus. 
The study contains three sessions: tutorial, drawing, and interview. 
The entire study took about two hours per participant.
\item[Participants] We recruited 12 participants (seven males and five females).
All participants are non-professional artists, and only 3 of them have experience in physical-painting or digital-painting.
\cgfhl{
We ask each participant to draw one of the six clipart (three chairs and three airplanes).
}
\item[Tutorial session] We designed the tutorial session to help the participant familiarize with the drawing interface we provided and different functions in our drawing interface.
There are four stages of the tutorial session that aim to let the user familiarize with functions, including (i) drawing single curves, (ii) connecting curves, (iii) filling color, (iv) how to draw primitives, and (v) the layering concept.
For each function, we prepared target results (e.g., rectangles with different sizes or specific layering) and asked them to use the tools to match the targets.
In total, we allow the participants to practice in the tutorial session for at most 15 minutes.
\item[Drawing session] In the drawing session, we show the input clipart to the participants and ask them to design clipart of the same object but from three other viewpoints.
For chair clipart, the input viewpoint is from the front viewpoint, and then we ask the user to draw from the top, and the side, and the upper $45^{\circ}$ viewpoints.
For airplane clipart, the input viewpoint is from the top viewpoint, so we ask the user to draw the front, and the side, and the upper $45^{\circ}$ viewpoints.
Each participant design the clipart with and without our visual scaffold \minorhl{in a random order to avoid bias.}
\item[Interview] In the end, we collect feedback from each participant on different aspects of the assistive drawing interface.
In the questionnaire, we asked participants about (i) their drawing experience, (ii) their thoughts on difficulties of drawing different viewpoints, and (iii) how they think the visual scaffold aids them during the drawing session.
In the last part, we asked the following questions:
\begin{itemize}
    \item do they think the provided visual scaffold is helpful?
    \item how the provided visual scaffold affect them while drawing?
    \item other suggestions or comments on how they think the provided visual scaffold aid can be improved.
\end{itemize}
Please find the full version of our questionnaire in \secname~3 of the supplemental material.
\item[Drawing result] We show several clipart designed by the participants in \figname~\ref{fig:user_result}.
\minorhll{
Noted that we do not have the exact correct design for all the clipart we asked the participants to design.
Thus, in the following discussion, we use the artist's results as the best reference design since the artists are usually more familiar with the workflow and are more likely to design better results.
In \mbox{\figname~\ref{fig:user_result}}, we can observe that the participant can draw a better ratio between the width and the height of the purple sofa case.
The design result with the visual scaffold is closer to the artist's result than the result without the visual scaffold.
For the purple sofa case, the participant can design better clipart from the upper $45^{\circ}$ by mostly followed the provided visual scaffold because the generated 3D shape matched the expected underlying geometry.
Simultaneously, even though sometimes the visual scaffold's geometry is not accurate at every location, it can still assist the participants in designing the clipart.
For the blue airplane case (shown in the last row in \mbox{\figname~\ref{fig:user_result})}, the participant can place the wing at the position closer to the artist's result by designing with the provided visual scaffold.
}
Please see the complete results from all participants in the supplemental material.
\end{description}
\subsubsection{Interview result and discussion}
\begin{description}[style=unboxed,leftmargin=0cm]
\item[Viewpoint Survey]
\cgfhl{
We asked the participants about their thoughts on difficulties of drawing different viewpoints.
All of the participants agree that the difficulties vary across viewpoints. 
Among all participants, eleven out of twelve consider the upper $45^{\circ}$ viewpoint more challenging to draw than the rest viewpoints.
More specifically, for airplane clipart, $83.3\%$ of participants thought the upper $45^{\circ}$ viewpoint is the hardest to draw, and $16.7\%$ of them thought the side viewpoint is the hardest one.
In terms of chair clipart, all participants agree that the upper $45^{\circ}$ viewpoint is the hardest to draw.
}
\item[Scaffold aids survey]
\cgfhl{
Eleven of twelve participants ($91.7\%$) considered the provided visual scaffold is helpful during the drawing process.
And they thought the visual scaffold is helpful in the following ways:
}
\begin{itemize}
\item \cgfhl{two participants thought the visual scaffold helps them see the overview at the beginning of the design process. It gives them a better idea of how to draw the different layers composing the target object.}
\item \cgfhl{five participants thought the visual scaffold helps them estimate the ratio and scale between different parts of the target object.}
\item \cgfhl{two participants thought the visual scaffold reminds them some details they ignore}
\item \cgfhl{two participants thought the visual scaffold helps them to arrange the layering while designing the target object.}
\end{itemize}
\end{description}
\cgfhl{
According to the above user feedback, we can observe that our visual scaffold helps design clipart under an unseen viewpoint, especially useful for estimating the ratio and scale between parts.
Meanwhile, two participants thought the visual scaffold might interfere with their imagination, thus hindering the design process.
The potential reason is that the visual scaffold is sometimes inaccurate, which might conflict with the participants' imagination of the target object.
}

\begin{figure}[t!]
\centering
\includegraphics[width=\linewidth]{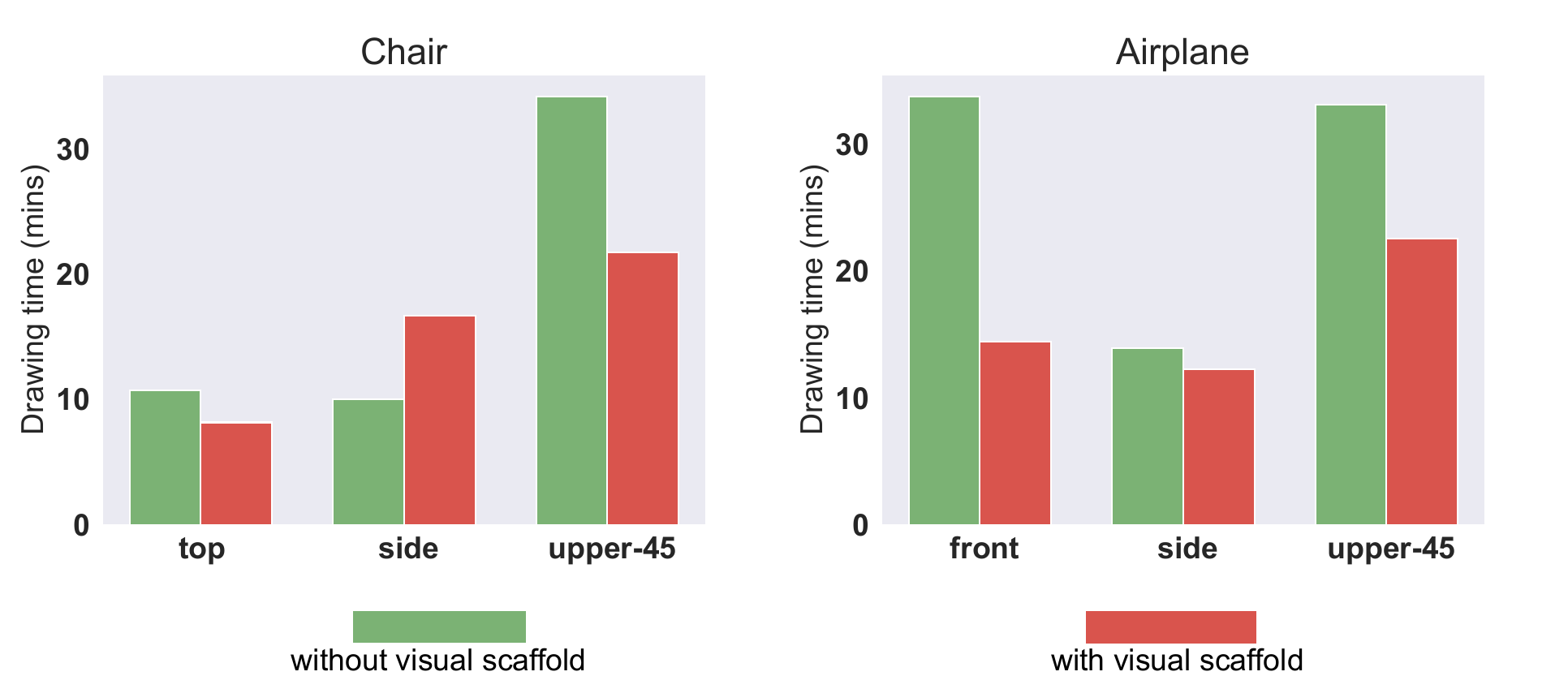}
\caption{
We show the drawing time statistics of the user study.
On average, the participants can save $15\%$ of drawing time for chair cases and $40\%$ for airplane cases.
The provided visual scaffold is most helpful in drawing the airplane from the front viewpoint (saved $57\%$ of drawing time).
And it is most helpful for drawing both the chair and the airplane from the upper $45^{\circ}$ viewpoint. 
}
\label{fig:draw_time}
\end{figure}

Overall, most of the participants can design the clipart from the target viewpoint in shorter periods.
As shown in \figname~\ref{fig:draw_time}, the participants saved time on drawing upper $45^{\circ}$ for both chair and airplane clipart.
With the provided visual scaffold, the participants can save $57\%$ of time when drawing complicated parts such as the airplane from the front viewpoint.
Meanwhile, we observed that the participants spent more time drawing the chair from the side viewpoint.
From the user feedback, we found out that this is because the visual scaffold provides additional details from the side viewpoint compared to their imagination.
Even though they spent more time drawing with visual scaffold from the side viewpoint, they can draw more details, leading to better results.

%% file: conclusion.tex
\section{Conclusion}
In this paper, we propose an assistive system that aids the users in designing clipart from unseen viewpoints.
The core of our system is a user-assisted curve extrusion method that combines the user-provided structural annotation and a guiding 3D mesh that is synthesized by the single-view shape reconstruction method.
We render the generated 3D shape using the same style of input clipart into the visual scaffold.
We conducted a user study with 12 users.
With our visual scaffold, we found out that the users can design better quality clipart from unseen viewpoints using a shorter time than designing without the provided visual scaffold.

\begin{figure}[t!]
\centering
\includegraphics[width=\linewidth]{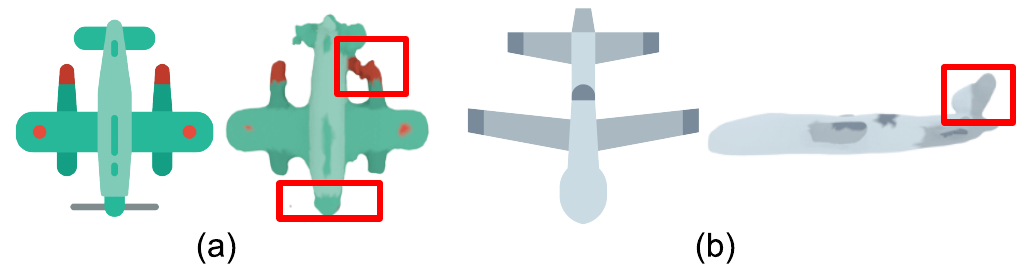}
\caption{
\minorhl{
We show two examples of problematic guiding shapes reconstructed using a learning-based 3D reconstruction method.
Given the input clipart on the left in both (a) and (b), we can observe the broken reconstruction results highlighted in red rectangles.
The reconstruction method fails to reconstruct some thin structures (e.g.,~propeller in (a)) or add some unnecessary structures (e.g.,~the vertical stabilizer in (b)).
}
}
\label{fig:failure}
\end{figure}

\cgfhl{
There are two parts of limitation in our proposed system, \ie~the 3D reconstruction part, and the user-assisted design part.
The limitations of our 3D reconstruction method are twofold. 
\minorhl{
The first part is because we choose to use a pre-trained model of an existing learning-based 3D reconstruction method~\mbox{\cite{lun20173d}} to reconstruct the guiding mesh, the quality of predicted shapes is usually not satisfying. 
Besides the reason mentioned in \mbox{\secname~\ref{sec:singleview}}, another reason is that many clipart shapes are quite distinct from the training data.
For example, we can not find similar shapes in the training data for the two airplanes shown in \mbox{\figname~\ref{fig:failure}}; thus, the reconstructed guiding meshes are pretty noisy and broken.
Our current method is not able to recover the missing parts in the guiding mesh.
}
The second one is that our extrusion method can not model curvature along the extrusion direction.
Since the current extrusion method only extrudes the geometry with a single thickness value.
We plan to use different geometry representation to address this issue, such as voxels, bevels, or \mbox{geometry profiles used in~\cite{kelly2017bigsur}}.
}
\cgfhl{
In terms of the limitation of the user-assisted design part, we observed that designs clipart from a non-canonical viewpoint (\eg~the upper $45^{\circ}$ viewpoint) is much harder than a canonical viewpoint.
However, our current system does not provide step-by-step assistance for users.
In the future, we would love to explore how to design more intuitive tools, \eg~adding block-in guides, or other novel assistance to better aids the users in designing clipart from non-canonical viewpoints.
}

%% file: ack.tex
\section{Acknowledgement}
This work was supported in part by the Ministry of Science and Technology, Taiwan, under Grant MOST109-2634-F-002-032, 109-2218-E-002-026, and National Taiwan University.
And we are grateful to the National Center for High-performance Computing.
We want to thank Sheng-Jie Luo, Chi-Lan Yang, anonymous reviewers for insightful suggestions, and Seraphina Yong for proofreading parts of the paper.
I-Chao Shen was supported by the MediaTek Fellowship.

%% file: paper.bbl
\newcommand{\etalchar}[1]{$^{#1}$}
\begin{thebibliography}{\uppercase{MWYG20}}

\bibitem[Ado20]{illustrator}
\textsc{Adobe}:
\newblock Adobe illustrator 2020: Image trace, 2020.
\newblock URL: \url{http://www.adobe.com/}.

\bibitem[BDS{\etalchar{*}}12]{bouaziz2012shape}
\textsc{Bouaziz S., Deuss M., Schwartzburg Y., Weise T., Pauly M.}:
\newblock Shape-up: Shaping discrete geometry with projections.
\newblock In \emph{Computer Graphics Forum} (2012), vol.~31, Wiley Online
  Library, pp.~1657--1667.

\bibitem[BK04]{boykov2004experimental}
\textsc{Boykov Y., Kolmogorov V.}:
\newblock An experimental comparison of min-cut/max-flow algorithms for energy
  minimization in vision.
\newblock \emph{IEEE transactions on pattern analysis and machine intelligence
  26}, 9 (2004), 1124--1137.

\bibitem[BMP01]{belongie2001shape}
\textsc{Belongie S., Malik J., Puzicha J.}:
\newblock Shape context: A new descriptor for shape matching and object
  recognition.
\newblock In \emph{Advances in neural information processing systems} (2001),
  pp.~831--837.

\bibitem[CFG{\etalchar{*}}15]{chang2015shapenet}
\textsc{Chang A.~X., Funkhouser T., Guibas L., Hanrahan P., Huang Q., Li Z.,
  Savarese S., Savva M., Song S., Su H., et~al.}:
\newblock Shapenet: An information-rich 3d model repository.
\newblock \emph{arXiv preprint arXiv:1512.03012} (2015).

\bibitem[CGW{\etalchar{*}}14]{Chen:2014:HAV}
\textsc{Chen H.-T., Grossman T., Wei L.-Y., Schmidt R.~M., Hartmann B.,
  Fitzmaurice G., Agrawala M.}:
\newblock History assisted view authoring for 3d models.
\newblock In \emph{Proceedings of the SIGCHI Conference on Human Factors in
  Computing Systems} (New York, NY, USA, 2014), CHI '14, ACM, pp.~2027--2036.
\newblock URL: \url{http://doi.acm.org/10.1145/2556288.2557009}, \href
  {https://doi.org/10.1145/2556288.2557009}
  {\path{doi:10.1145/2556288.2557009}}.

\bibitem[CSH19]{chen2019monocular}
\textsc{Chen X., Song J., Hilliges O.}:
\newblock Monocular neural image based rendering with continuous view control.
\newblock In \emph{Proceedings of the IEEE International Conference on Computer
  Vision} (2019), pp.~4090--4100.

\bibitem[CXG{\etalchar{*}}16]{choy20163d}
\textsc{Choy C.~B., Xu D., Gwak J., Chen K., Savarese S.}:
\newblock 3d-r2n2: A unified approach for single and multi-view 3d object
  reconstruction.
\newblock In \emph{European conference on computer vision} (2016), Springer,
  pp.~628--644.

\bibitem[DFRS03]{DeCarlo:2003:SCF}
\textsc{DeCarlo D., Finkelstein A., Rusinkiewicz S., Santella A.}:
\newblock Suggestive contours for conveying shape.
\newblock \emph{ACM Transactions on Graphics (Proc. SIGGRAPH) 22}, 3 (July
  2003), 848--855.

\bibitem[DN19]{Dai_2019_CVPR}
\textsc{Dai A., Niessner M.}:
\newblock Scan2mesh: From unstructured range scans to 3d meshes.
\newblock In \emph{The IEEE Conference on Computer Vision and Pattern
  Recognition (CVPR)} (June 2019).

\bibitem[DSTB16]{dosovitskiy2016learning}
\textsc{Dosovitskiy A., Springenberg J.~T., Tatarchenko M., Brox T.}:
\newblock Learning to generate chairs, tables and cars with convolutional
  networks.
\newblock \emph{IEEE transactions on pattern analysis and machine intelligence
  39}, 4 (2016), 692--705.

\bibitem[DTM96]{debevec1996modeling}
\textsc{Debevec P.~E., Taylor C.~J., Malik J.}:
\newblock Modeling and rendering architecture from photographs: A hybrid
  geometry-and image-based approach.
\newblock In \emph{Proceedings of the 23rd annual conference on Computer
  graphics and interactive techniques} (1996), pp.~11--20.

\bibitem[FSG17]{fan2017point}
\textsc{Fan H., Su H., Guibas L.~J.}:
\newblock A point set generation network for 3d object reconstruction from a
  single image.
\newblock In \emph{Proceedings of the IEEE conference on computer vision and
  pattern recognition} (2017), pp.~605--613.

\bibitem[GFK{\etalchar{*}}18]{groueix2018}
\textsc{Groueix T., Fisher M., Kim V.~G., Russell B., Aubry M.}:
\newblock {AtlasNet: A Papier-M\^ach\'e Approach to Learning 3D Surface
  Generation}.
\newblock In \emph{Proceedings IEEE Conf. on Computer Vision and Pattern
  Recognition (CVPR)} (2018).

\bibitem[HDS{\etalchar{*}}18]{hoshyari2018perception}
\textsc{Hoshyari S., Dominici E.~A., Sheffer A., Carr N., Wang Z., Ceylan D.,
  Shen I., et~al.}:
\newblock Perception-driven semi-structured boundary vectorization.
\newblock \emph{ACM Transactions on Graphics (TOG) 37}, 4 (2018), 118.

\bibitem[HLW{\etalchar{*}}17]{how2sketch}
\textsc{Hennessey J.~W., Liu H., Winnemöller H., Dontcheva M., Mitra N.~J.}:
\newblock How2sketch: Generating easy-to-follow tutorials for sketching 3d
  objects.
\newblock \emph{Symposium on Interactive 3D Graphics and Games} (2017).

\bibitem[KBH06]{Kaz06Poisson}
\textsc{Kazhdan M., Bolitho M., Hoppe H.}:
\newblock Poisson surface reconstruction.
\newblock In \emph{Proceedings of the Fourth Eurographics Symposium on Geometry
  Processing} (Goslar, DEU, 2006), SGP ’06, Eurographics Association,
  p.~61–70.

\bibitem[KFWM17]{kelly2017bigsur}
\textsc{Kelly T., Femiani J., Wonka P., Mitra N.~J.}:
\newblock Bigsur: large-scale structured urban reconstruction.
\newblock \emph{ACM Transactions on Graphics 36}, 6 (2017).

\bibitem[KL11]{kopf2011depixelizing}
\textsc{Kopf J., Lischinski D.}:
\newblock Depixelizing pixel art.
\newblock \emph{ACM Transactions on graphics (TOG) 30}, 4 (2011), 99.

\bibitem[KLA19]{karras2019style}
\textsc{Karras T., Laine S., Aila T.}:
\newblock A style-based generator architecture for generative adversarial
  networks.
\newblock In \emph{Proceedings of the IEEE Conference on Computer Vision and
  Pattern Recognition} (2019), pp.~4401--4410.

\bibitem[KLS{\etalchar{*}}13]{Kopf2013}
\textsc{Kopf J., Langguth F., Scharstein D., Szeliski R., Goesele M.}:
\newblock Image-based rendering in the gradient domain.
\newblock \emph{ACM Transactions on Graphics (Proceedings of SIGGRAPH Asia
  2013) 32}, 6 (2013).

\bibitem[LALR16]{Liu:2016:DI}
\textsc{Liu Y., Agarwala A., Lu J., Rusinkiewicz S.}:
\newblock Data-driven iconification.
\newblock In \emph{International Symposium on Non-Photorealistic Animation and
  Rendering (NPAR)} (May 2016).

\bibitem[LGK{\etalchar{*}}17]{lun20173d}
\textsc{Lun Z., Gadelha M., Kalogerakis E., Maji S., Wang R.}:
\newblock 3d shape reconstruction from sketches via multi-view convolutional
  networks.
\newblock In \emph{2017 International Conference on 3D Vision (3DV)} (2017),
  IEEE, pp.~67--77.

\bibitem[LHES19a]{Lopes_2019_ICCV}
\textsc{Lopes R.~G., Ha D., Eck D., Shlens J.}:
\newblock A learned representation for scalable vector graphics.
\newblock In \emph{The IEEE International Conference on Computer Vision (ICCV)}
  (October 2019).

\bibitem[LHES19b]{lopes2019learned}
\textsc{Lopes R.~G., Ha D., Eck D., Shlens J.}:
\newblock A learned representation for scalable vector graphics.
\newblock In \emph{Proceedings of the IEEE International Conference on Computer
  Vision} (2019), pp.~7930--7939.

\bibitem[lJ19]{liu2019cubic}
\textsc{liu H.-T.~D., Jacobson A.}:
\newblock Cubic stylization.
\newblock \emph{ACM Transactions on Graphics (TOG) 38}, 6 (2019), 1--10.

\bibitem[LKS15]{Lun:2015:StyleSimilarity}
\textsc{Lun Z., Kalogerakis E., Sheffer A.}:
\newblock Elements of style: Learning perceptual shape style similarity.
\newblock \emph{ACM Transactions on Graphics 34}, 4 (2015).

\bibitem[LKWS16]{Lun:2016:StyleTransfer}
\textsc{Lun Z., Kalogerakis E., Wang R., Sheffer A.}:
\newblock Functionality preserving shape style transfer.
\newblock \emph{ACM Transactions on Graphics 35}, 6 (2016).

\bibitem[LLCL19]{liu2019soft}
\textsc{Liu S., Li T., Chen W., Li H.}:
\newblock Soft rasterizer: A differentiable renderer for image-based 3d
  reasoning.
\newblock In \emph{Proceedings of the IEEE International Conference on Computer
  Vision} (2019), pp.~7708--7717.

\bibitem[LPL{\etalchar{*}}18]{li2018robust}
\textsc{Li C., Pan H., Liu Y., Tong X., Sheffer A., Wang W.}:
\newblock Robust flow-guided neural prediction for sketch-based freeform
  surface modeling.
\newblock \emph{ACM Transactions on Graphics (TOG) 37}, 6 (2018), 1--12.

\bibitem[LYH{\etalchar{*}}15]{Luo:2015:LOL}
\textsc{Luo S.-J., Yue Y., Huang C.-K., Chung Y.-H., Imai S., Nishita T., Chen
  B.-Y.}:
\newblock Legolization: Optimizing lego designs.
\newblock \emph{ACM Transactions on Graphics (Proc. SIGGRAPH Asia 2015) 34}, 6
  (2015), 222:1--222:12.

\bibitem[LZC11]{Lee:2011:SRU}
\textsc{Lee Y.~J., Zitnick C.~L., Cohen M.~F.}:
\newblock Shadowdraw: Real-time user guidance for freehand drawing.
\newblock \emph{ACM Trans. Graph. 30}, 4 (July 2011), 27:1--27:10.
\newblock URL: \url{http://doi.acm.org/10.1145/2010324.1964922}, \href
  {https://doi.org/10.1145/2010324.1964922}
  {\path{doi:10.1145/2010324.1964922}}.

\bibitem[LZX{\etalchar{*}}08]{liu2008local}
\textsc{Liu L., Zhang L., Xu Y., Gotsman C., Gortler S.~J.}:
\newblock A local/global approach to mesh parameterization.
\newblock In \emph{Computer Graphics Forum} (2008), vol.~27, Wiley Online
  Library, pp.~1495--1504.

\bibitem[MWYG20]{mo2020pt2pc}
\textsc{Mo K., Wang H., Yan X., Guibas L.}:
\newblock {PT2PC}: Learning to generate 3d point cloud shapes from part tree
  conditions.
\newblock \emph{arXiv preprint arXiv:2003.08624} (2020).

\bibitem[MZL{\etalchar{*}}09]{mzlsgm_abstraction_siga_09}
\textsc{Mehra R., Zhou Q., Long J., Sheffer A., Gooch A., Mitra N.~J.}:
\newblock Abstraction of man-made shapes.
\newblock \emph{{ACM} Transactions on Graphics 28}, 5 (2009), \#137, 1--10.

\bibitem[NMOG19]{niemeyer2019differentiable}
\textsc{Niemeyer M., Mescheder L., Oechsle M., Geiger A.}:
\newblock Differentiable volumetric rendering: Learning implicit 3d
  representations without 3d supervision.
\newblock \emph{arXiv preprint arXiv:1912.07372} (2019).

\bibitem[NPLT{\etalchar{*}}19]{nguyen2019hologan}
\textsc{Nguyen-Phuoc T., Li C., Theis L., Richardt C., Yang Y.-L.}:
\newblock Hologan: Unsupervised learning of 3d representations from natural
  images.
\newblock In \emph{Proceedings of the IEEE International Conference on Computer
  Vision} (2019), pp.~7588--7597.

\bibitem[OTW{\etalchar{*}}19]{olszewski2019transformable}
\textsc{Olszewski K., Tulyakov S., Woodford O., Li H., Luo L.}:
\newblock Transformable bottleneck networks.
\newblock In \emph{Proceedings of the IEEE International Conference on Computer
  Vision} (2019), pp.~7648--7657.

\bibitem[PXW18]{Peng:2018:AS}
\textsc{Peng M., Xing J., Wei L.-Y.}:
\newblock Autocomplete 3d sculpting.
\newblock \emph{ACM Trans. Graph. 37}, 4 (July 2018), 132:1--132:15.
\newblock URL: \url{http://doi.acm.org/10.1145/3197517.3201297}, \href
  {https://doi.org/10.1145/3197517.3201297}
  {\path{doi:10.1145/3197517.3201297}}.

\bibitem[PYY{\etalchar{*}}17]{park2017transformation}
\textsc{Park E., Yang J., Yumer E., Ceylan D., Berg A.~C.}:
\newblock Transformation-grounded image generation network for novel 3d view
  synthesis.
\newblock In \emph{Proceedings of the ieee conference on computer vision and
  pattern recognition} (2017), pp.~3500--3509.

\bibitem[PZ17]{Soft3DReconstruction}
\textsc{Penner E., Zhang L.}:
\newblock Soft 3d reconstruction for view synthesis.

\bibitem[SBS19]{DeepSketch}
\textsc{Smirnov D., Bessmeltsev M., Solomon J.}:
\newblock Deep sketch-based modeling of man-made shapes.
\newblock \emph{CoRR abs/1906.12337} (2019).
\newblock URL: \url{http://arxiv.org/abs/1906.12337}, \href
  {http://arxiv.org/abs/1906.12337} {\path{arXiv:1906.12337}}.

\bibitem[SCD{\etalchar{*}}06]{seitz2006comparison}
\textsc{Seitz S.~M., Curless B., Diebel J., Scharstein D., Szeliski R.}:
\newblock A comparison and evaluation of multi-view stereo reconstruction
  algorithms.
\newblock In \emph{2006 IEEE computer society conference on computer vision and
  pattern recognition (CVPR'06)} (2006), vol.~1, IEEE, pp.~519--528.

\bibitem[SHL{\etalchar{*}}18]{sun2018multiview}
\textsc{Sun S.-H., Huh M., Liao Y.-H., Zhang N., Lim J.~J.}:
\newblock Multi-view to novel view: Synthesizing novel views with self-learned
  confidence.
\newblock In \emph{European Conference on Computer Vision} (2018).

\bibitem[SIJ{\etalchar{*}}07]{SIJSW07}
\textsc{Schmidt R., Isenberg T., Jepp P., Singh K., Wyvill B.}:
\newblock Sketching, scaffolding, and inking: A visual history for interactive
  3d modeling.
\newblock In \emph{NPAR '07: Proceedings of the 5th international symposium on
  Non-photorealistic animation and rendering} (2007), pp.~23--32.
\newblock URL:
  \url{http://www.dgp.toronto.edu/~rms/pubs/ScaffoldingNPAR07.html}.

\bibitem[SKSK09]{SKSK09}
\textsc{Schmidt R., Khan A., Singh K., Kurtenbach G.}:
\newblock Analytic drawing of 3d scaffolds.
\newblock \emph{ACM Transactions on Graphics 28}, 5 (2009).
\newblock Proceedings of SIGGRAPH ASIA 2009.
\newblock URL: \url{http://www.dgp.toronto.edu/~rms/pubs/DrawingSGA09.html}.

\bibitem[SLHC12]{Shen:2012:SDM}
\textsc{Shen L.-T., Luo S.-J., Huang C.-K., Chen B.-Y.}:
\newblock Sd models: Super-deformed character models.
\newblock \emph{Computer Graphics Forum 31}, 7 (2012), 2067--2075.
\newblock (Pacific Graphics 2012 Conference Proceedings).

\bibitem[Ste03]{hqx}
\textsc{Stepin M.}:
\newblock Hqx, 2003.
\newblock URL:
  \url{http://web.archive.org/web/20070717064839/www.hiend3d.com/hq4x.html}.

\bibitem[TDB16]{MV3SI}
\textsc{Tatarchenko M., Dosovitskiy A., Brox T.}:
\newblock Multi-view 3d models from single images with a convolutional network.
\newblock In \emph{European Conference on Computer Vision (ECCV)} (2016).

\bibitem[TZEM17]{drcTulsiani17}
\textsc{Tulsiani S., Zhou T., Efros A.~A., Malik J.}:
\newblock Multi-view supervision for single-view reconstruction via
  differentiable ray consistency.
\newblock In \emph{Computer Vision and Pattern Regognition (CVPR)} (2017).

\bibitem[{Vec}20]{vector_magic}
\textsc{{Vector Magic}}:
\newblock Cedar lake ventures, 2020.
\newblock URL: \url{https://vectormagic.com/}.

\bibitem[VK78]{vandenberg1978mental}
\textsc{Vandenberg S.~G., Kuse A.~R.}:
\newblock Mental rotations, a group test of three-dimensional spatial
  visualization.
\newblock \emph{Perceptual and motor skills 47}, 2 (1978), 599--604.

\bibitem[XCW14]{Xing:2014:APR}
\textsc{Xing J., Chen H.-T., Wei L.-Y.}:
\newblock Autocomplete painting repetitions.
\newblock \emph{ACM Trans. Graph. 33}, 6 (Nov. 2014), 172:1--172:11.
\newblock URL: \url{http://doi.acm.org/10.1145/2661229.2661247}, \href
  {https://doi.org/10.1145/2661229.2661247}
  {\path{doi:10.1145/2661229.2661247}}.

\bibitem[XKG{\etalchar{*}}16]{Xing:2016:EIT}
\textsc{Xing J., Kazi R.~H., Grossman T., Wei L.-Y., Stam J., Fitzmaurice G.}:
\newblock Energy-brushes: Interactive tools for illustrating stylized elemental
  dynamics.
\newblock In \emph{Proceedings of the 29th Annual Symposium on User Interface
  Software and Technology} (New York, NY, USA, 2016), UIST '16, ACM,
  pp.~755--766.
\newblock URL: \url{http://doi.acm.org/10.1145/2984511.2984585}, \href
  {https://doi.org/10.1145/2984511.2984585}
  {\path{doi:10.1145/2984511.2984585}}.

\bibitem[YK12]{Yumer:2012:COA}
\textsc{Yumer M.~E., Kara L.~B.}:
\newblock Co-abstraction of shape collections.
\newblock \emph{ACM Trans. Graph. 31}, 6 (Nov. 2012).
\newblock URL: \url{https://doi.org/10.1145/2366145.2366185}, \href
  {https://doi.org/10.1145/2366145.2366185}
  {\path{doi:10.1145/2366145.2366185}}.

\bibitem[ZTS{\etalchar{*}}16]{zhou2016view}
\textsc{Zhou T., Tulsiani S., Sun W., Malik J., Efros A.~A.}:
\newblock View synthesis by appearance flow.
\newblock In \emph{European Conference on Computer Vision} (2016).

\end{thebibliography}
